%% file: facos.tex
\documentclass[journal]{IEEEtran}
\pdfoutput=1
\usepackage{dashrule}
\usepackage{algorithm}
\usepackage{algorithmic}
\usepackage{xcolor}
\usepackage{graphicx}
\usepackage{tikz}
\usepackage{afterpage}
\usepackage{color}
\usepackage{amsmath}
\usepackage{float}
\usepackage[numbers]{natbib}
\usepackage{tikz}
\usetikzlibrary{shapes}
\usetikzlibrary{matrix}
\usetikzlibrary{patterns}
\usetikzlibrary{backgrounds}
\usepackage{pgfplots}
\pgfplotsset{compat=1.3}
\usetikzlibrary{scopes}
\usepackage{arydshln}
\usepackage{comment}
\usepackage{subcaption}
\usepackage{amssymb}
\usepackage{wasysym}
\usepackage{graphicx}
\usepackage{multirow}

\usepackage[colorlinks=true, allcolors=blue]{hyperref}
\newcommand{\heading}[1]{{\vspace{0.4pt}\noindent{\textbf{#1}}}}

\newcommand{\circlednumber}[1]{%
  \tikz[baseline=(char.base)]{
    \node[shape=circle, draw, fill=black, text=white, inner sep=0.2pt] (char) {\textbf{#1}};
  }%
}

\begin{document}

\title{FACOS: Enabling Privacy Protection Through Fine-Grained Access Control with On-chain and Off-chain System}

\author{Chao~Liu,~\IEEEmembership{Member,~IEEE,}
        Cankun~Hou,~\IEEEmembership{Student Member,~IEEE,}
        Tianyu~Jiang,~\IEEEmembership{Student Member,~IEEE,}
        Jianting Ning,~\IEEEmembership{Member,~IEEE,}
        Hui~Qiao,
        Yusen~Wu, ~\IEEEmembership{Member,~IEEE}

    \thanks{C. Liu and C. Hou contributed equally to this work.}
     \thanks{Corresponding author: Jianting Ning (e-mail: jtning88@gmail.com), Hui Qiao and Yusen Wu.}
    \thanks{C. Hou and T. Jiang were with Fujian Provincial Key Laboratory of Data Intensive Computing and the Key Laboratory of Intelligent Computing and Information Processing, Fujian, China} 
    \thanks{H. Qiao was with the  School of Computer Science and Technology, Xidian University, Xi'an Shaanxi 710071, China.} 
    \thanks{Y. Wu was with the Department of Computer Science, University of Miami, Miami, FL, 33146 USA. }
     \thanks{J. Ning is with the Key Laboratory of Analytical Mathematics and Applications (Ministry of Education), College of Computer and Cyber Security, Fujian Normal University, Fuzhou 350007, China, and also with the Faculty of Data Science, City University of Macau, Macau, China.} 
}


\maketitle

\begin{abstract}
Data-driven landscape across finance, government, and healthcare, the continuous generation of information demands robust solutions for secure storage, efficient dissemination, and fine-grained access control. Blockchain technology emerges as a significant tool, offering decentralized storage while upholding the tenets of data security and accessibility. However, on-chain and off-chain strategies are still confronted with issues such as untrusted off-chain data storage, absence of data ownership, limited access control policy for clients, and a deficiency in data privacy and auditability. To solve these challenges, we propose a permissioned blockchain-based privacy-preserving fine-grained access control on-chain and off-chain system, namely FACOS. We applied three fine-grained access control solutions and comprehensively analyzed them in different aspects, which provides an intuitive perspective for system designers and clients to choose the appropriate access control method for their systems. Compared to similar work that only stores encrypted data in centralized or non-fault-tolerant IPFS systems, we enhanced off-chain data storage security and robustness by utilizing a highly efficient and secure asynchronous Byzantine fault tolerance (BFT) protocol in the off-chain environment. As each of the clients needs to be verified and authorized before accessing the data, we involved the Trusted Execution Environment (TEE)-based solution to verify the credentials of clients.
Additionally, our evaluation results demonstrated that our system\footnote{https://github.com/cliu717/AsynchronousStorage} offers better scalability and practicality than other state-of-the-art designs. We deployed our system on Alibaba Cloud and Tencent Cloud and conducted multiple evaluations. The results indicate that it takes about 2.79 seconds for a client to execute the protocol for uploading and about 0.96 seconds for downloading. Compared to other decentralized systems, our system exhibits efficient latency for both download and upload operations.
\end{abstract}


\input{tex/introduction}
\input{tex/relatework}
\input{tex/preliminaries}
\input{tex/securitymodel}
\input{tex/system}
\input{tex/security}

\input{tex/evaluation}
\input{tex/conclusion}

\bibliographystyle{abbrv}
\bibliography{facos}

\end{document}

%% file: tex/introduction.tex
\section{Introduction}
\begin{table*}[!ht]
	\centering
	\scriptsize
	\begin{tabular}{|c|c|c|c|c|c|c|c|}
		\hline
		Protocol  &\!\!\!\!\!\!\!\! \begin{tabular}{c}Crypto\\ operation\end{tabular}\!\!\!\!\!\!\!\! & \!\!\!\!\begin{tabular}{c}Arbitrary\\ attribute\end{tabular}\!\!\!\! &\!\!\!\!\!\!\!\!\begin{tabular}{c}Full\\ security\end{tabular}\!\!\!\!\!\!\!\! 
       &\!\!\!\!\!\!\!\!\begin{tabular}{c}Unrestricted\\ policies\end{tabular}\!\!\!\!\!\!\!\!
       &\!\!\!\!\!\!\!\!\begin{tabular}{c}Standard\\ assumption\end{tabular}\!\!\!\!\!\!\!\!
       &\!\!\!\!\!\!\!\!\begin{tabular}{c}Third \\ party\end{tabular}\!\!\!\!\!\!\!\!
       &\!\!\!\!\!\!\!\!\begin{tabular}{c}Fault\\ tolerant\end{tabular}\!\!\!\!\!\!\!\!\\
		\hline
		ABE~\cite{agrawal2017fame} & pairing-based & \CIRCLE & \CIRCLE  &\CIRCLE &\CIRCLE&\CIRCLE&\Circle\\
		\hdashline[1pt/1pt]
        BE~\cite{fiat1994broadcast} & symmetric & \Circle  & \Circle  &\Circle &\Circle&\Circle&\Circle\\
		\hdashline[1pt/1pt]
         TE~\cite{shoup1998securing} & elliptic curve & \CIRCLE & \Circle &\CIRCLE &\Circle&\CIRCLE&\CIRCLE \\
		\hline
	\end{tabular} \
	\vspace{3pt}
	\caption{THE THREE ACCESS CONTROL METHODS. \CIRCLE~DENOTES FULL SUPPORT.} 
	\label{tab:detailedthreemethods}
\vspace{-6pt}
\end{table*}

\IEEEPARstart{I}{n} the fields of finance, government, and healthcare, amounts of data are generated daily, necessitating secure storage, efficient dissemination, and easy access. Financial transactions are generated with every payment, administrative processes continuously produce government records, and the medical field constantly generates data through procedures like CT-SCANs, X-rays, and prescriptions. The utmost priority is to ensure the confidentiality and accessibility of these sensitive data through an immutable and private system. Blockchain technology \cite{belchior2021survey, li2023designated} emerges as an ideal solution, offering decentralized storage \cite{benisi2020blockchain} for diverse users and organizations. Within a blockchain network, government records, for instance, can be effortlessly shared among trusted peers such as government agencies and officials, while preserving data security, integrity, and consistency. This technology's foundation rests on sequentially connected blocks, secured by cryptographic hashing, guaranteeing the immutability of transactions. It efficiently accommodates complicated record details, including historical and current information, making them readily accessible to authorized parties. To manage the sheer volume of records, a distributed on-chain storage system with a peer-to-peer structure \cite{wang2006structure} is imperative, ensuring the persistent and secure storage of data.

While blockchain technology holds great potential for addressing contemporary challenges in transaction management, applying traditional public blockchain designs to data storage and transfer presents inherent issues. Notably, the demands for privacy and security when dealing with sensitive data significantly diverge from those of public data. Furthermore, there's an imperative for a scalable, high-throughput system capable of efficiently handling substantial data volumes with minimal computational expenses to support real-world operations. It's crucial to acknowledge that conventional public blockchain features may not align with trusted data governance, which prioritizes privacy and efficiency. This misalignment leads to problems such as the absence of permission and access control management, the inefficiency of read and write, and low privacy for the data. Consequently, novel approaches to large sensitive data management using blockchain technology are essential.

In order to enhance the efficiency of data transmission, off-chain storage \cite{hepp2018chain} emerges as a vital solution, primarily propelled by fundamental concerns related to block size and system scalability (e.g., images are not suitable for storage on the blockchain because they typically have excessively large file sizes). We found various forms of sensitive data are bound by regulatory obligations mandating controlled and secure settings, such as compliance with data protection regulations. These requirements frequently clash with the idea of public storage, regardless of whether it is encrypted or not, thus making the adoption of \textit{trusted} off-chain storage solutions imperative. However, current off-chain solutions are unable to mitigate the problem of a single point of failure, which can lead to data loss and corruption, thereby impacting the accessibility and reliability of data access. 

Moreover, although on-chain storage may not be a practical choice for accommodating large datasets, the advantages of on-chain technology can still be effectively harnessed within off-chain storage solutions. One noteworthy approach involves storing the hash of a data fragment on the blockchain ledger. Given the relatively small size of a hash, the associated storage costs remain economical, allowing authorized users to query the actual dataset using the hash address securely and efficiently. The central challenge then centers on establishing a seamless connection between the hash recorded on the blockchain and the physical storage location of the data. 

We highlight our three access control solutions to enhance the understanding of fine-grained access control. In FACOS, fine-grained access control has three key implications. First, in contrast to most blockchain-based access control methods, which rely on attribute encryption, a comparison of these three access control methods reveals distinct pros and cons, as illustrated in Table \ref{tab:detailedthreemethods}. FACOS provides users with the flexibility to choose the most suitable algorithm based on their specific requirements and environmental settings. Second, both attribute encryption and threshold encryption methods enable users to access data with permissions in a more fine-grained manner, whether based on attributes or labels. Meanwhile, broadcast encryption allows groups with certain properties to access specific data. Our fine-grained access control can be defined as who can access the data, when the data can be accessed, and what attributes and labels can access the data, etc. Finally, it's worth noting that the performance of the system can be influenced by the use of smart contract-based fine-grained access control methods. In FACOS, we have chosen not to rely on smart contracts due to their potential impact on the overall system efficiency. We achieve greater system efficiency by directly integrating access control methods into the underlying protocols.
\begin{itemize}
    \item \textit{Attribute-based Encryptio (ABE)}: The advantages of ABE include the ability for data owners to determine who can access data based on a client's specific attributes with full security. ABE relies on a trusted third party to generate clients' public and private keys, and it also depends on complex cryptographic algorithms, which generally result in lower efficiency.
 
    \item \textit{Broadcast encryption (BE)}: Broadcast encryption can be used to disseminate encrypted information within a group that has access to this information and broadcast encryption algorithms exclusively utilize symmetric encryption methods, greatly enhancing encryption and decryption efficiency. When the number of clients in the entire group who cannot access this message is relatively small, the ciphertext length is smaller. Broadcast encryption can only provide access to a group of clients and cannot offer access control for clients with different attributes. Furthermore, when clients form a complete binary tree to facilitate efficient data distribution and access control, how to arrange them into an efficient full binary tree is crucial. 

    \item {Threshold encryption (TE)}: Threshold encryption, also known as distributed encryption, is designed to prevent single points of failure and ensure more effective data confidentiality. Each replica only uses its private key to generate a portion of the decryption ciphertext. The client can decrypt the data only when he/she receives a threshold number of decryption shares. Threshold encryption also relies on a centralized key generation center to generate private keys for all replicas and the system's public key. It's important to note that threshold encryption is generally slower during the decryption phase compared to the other two schemes. Chios \cite{duan2020intrusion} is the first work applied threshold encryption scheme in Pub/sub system.
\end{itemize}
\heading{Contributions.} To summarize, our contributions are as follows:
\begin{itemize}
    \item We have developed and implemented a client-based access control solution focusing on enhancing privacy and ensuring precise permission management, recognizing the diverse needs and roles of individual clients. Our approach encompasses three distinct access control methods for clients. We give a comprehensive analysis of these methods and this provides an intuitive view for the system designer and client to choose the appropriate one for their systems. Additionally, compared with the access control-based approach of using smart contracts, our fine-grained access control methods have high efficiency. 

    \item To enhance security, robustness, and intrusion tolerance on off-chain storage, we involve the significance of asynchronous Byzantine fault tolerance storage solutions. This novel and engineered solution recognizes the regulatory requirements for controlled and secure data settings, advocating for adopting off-chain storage to alleviate data transfer challenges.

    \item We introduce permissioned blockchain technology as a solution, providing various organizations with a secure, immutable, and confidential network for data storage. It underlines the importance of access control-based mechanisms to strengthen security and accountability. Meanwhile, we utilize the trusted TEE verifier to verify and authorize the clients to access the data.

    \item We conduct extensive evaluations to assess the efficiency of on-chain and off-chain solutions in our study. Our research includes empirical evidence and analysis that substantiates the effectiveness of these solutions in addressing data storage and transfer challenges. FACOS is open-sourced, which provides high scalability and practicality. 
 
\end{itemize}

%% file: tex/relatework.tex
\section{Related Work}
\subsection{On-chain and off-chain work}
\par The reliable interaction between on-chain and off-chain data forms a crucial cornerstone for the widespread adoption of blockchain systems. Addressing the expedient, efficient, and secure uploading of off-chain data onto the blockchain, while ensuring the atomicity, consistency, and security of both on-chain and off-chain data, stands as a pressing issue demanding resolution within the blockchain sphere. Chen \emph{et al.} \cite{chen2019toward} propose a quality-driven auction-based incentive mechanism based on a consortium blockchain that guarantees trust in both on-chain data and off-chain data. Cai \emph{et al.} \cite{cai2022scalable} propose a real-time, trusted data interactive, and fine-grained transaction supportable sharing framework, the core of which is a novel two-layer scaling blockchain design. 
In 2023, on the basis of constructing the off-chain transmission channel, Yu \emph{et al.} \cite{yu2023mehldt} propose a multielement hash lock data transfer mechanism (MeHLDT) for on-chain and off-chain.

\textit{However, existing solutions have not adequately addressed the issue of fault tolerance for off-chain data. This could lead to data loss, corruption, or errors during processing, thereby affecting the overall reliability and stability of the system.} 

\subsection{Fined-grained access control}
\par The main purpose of access control is to manage and control users' access to systems, resources, or data. Access controls restrict access to sensitive data, ensuring that only authorized users can view, modify, or share it. Furthermore, the distributed nature of blockchain helps reduce the risks of single points of failure and malicious behavior, thereby enhancing the overall security of the access control system. Attribute-based, broadcast encryption-based and threshold encryption-based access control approaches \cite{agrawal2017fame, bethencourt2007ciphertext, chen2015improved} are proposed to achieve fine-grained access control of data. Ren \emph{et al.} \cite{ren2021siledger} proposes an open, trusted, and decentralized access control mechanism based on blockchain and attribute-based encryption (ABE) in software-defined networks in Internet of Things (SDN-IoT) Networks. 
In \cite{mihaljevic2023approach}, a generic framework for conditional data access control within IoT is proposed. The generic framework is based on the employment of a dedicated secret key-based broadcast encryption scheme.
Yu \emph{et al.} \cite{yu2021blockchain} proposes a Shamir threshold cryptography scheme for the Industrial IoT (IIoT) data protection using blockchain. 

\textit{While the aforementioned approach excels in achieving secure data sharing, it does not adequately ensure the trustworthy storage of data. This limitation implies that despite safeguarding data during transmission and sharing, there might still exist potential threats to data during the storage phase, such as data tampering, loss, or other unauthorized alterations. }

\subsection{Diverse application scenarios}

\par In diverse application scenarios, the appropriateness of access control methods is contingent upon a multitude of factors, encompassing the nature of data, the intricacies of authorization prerequisites, the scale of user engagement, and more. Universality in access control solutions remains elusive due to the inherent intricacies of each distinct setting. 
Han \emph{et al.} \cite{han2023access} proposes an Internet of Things access control mechanism based on blockchain and inner product encryption, and employs blockchain technology to provide distributed and decentralized access control management in the Internet of Things.
Li \emph{et al.} \cite{li2023secure} presents a secure blockchain-assisted access control scheme for smart healthcare system in fog computing. All the operations of users are recorded on the blockchain by smart contract in order to ensure transparency and reliability of the system. 
Li \emph{et al.} \cite{li2023blockchain} presents Fabric-SCF.
The attribute-based access control model is deployed for access control, also utilizing smart contracts to define system processes and access policies to ensure the system's efficient operation.

\textit{However, most of the current solutions store and share raw data in plaintext on the blockchain, a common practice across various application domains. This approach can lead to the exposure of users' private data on the blockchain, potentially compromising their security and privacy.}

%% file: tex/preliminaries.tex
\section{Preliminaries}
 
\heading{Attribute-based Encryption}. 
Attribute-based cryptography  \cite{agrawal2017fame} is a widely employed encryption technique that enables fine-grained access control. The ABE system encompasses several key algorithms. The Setup Algorithm: This algorithm initializes the system's security parameter, denoted as $\lambda$, and generates both the public parameters $pk$ and a master key $msk$. Encryption Phase: Given the public key $pk$, an access structure (in Boolean expression form, e.g., ``attribute~1 AND attribute~2" or ``attribute~3 OR attribute~4"), and a plaintext $m$ as input, this process outputs a ciphertext $c$.Key Generation Phase: Given the master key 
$msk$ and a set of attributes (participating user's attributes) as input, this process outputs a secret key 
$sk$. Decryption Phase: Given the public key 
$pk$, ciphertext $c$, and secret key $sk$ as input, this process outputs a message $m$ (indicating decryption success, revealing the original message) or $\perp$ (indicating decryption failure).

\heading{Broadcast Encryption}.
The complete subtree broadcast encryption technique, devised by Naor, Naor, and Lotspiech (NNL) \cite{staddon2008content}, showcases efficiency particularly when the count of revoked recipients remains small. Consider a scenario with $N$ potential recipients. Picture a full binary tree with $N$ leaves, each signifying a recipient. Let us label the entire set of leaves as $N$. For any given node $l_i$ in the tree, $S_i$ is the set representing all leaves under the subtree rooted at $l_i$. This gives us a total of $2N-1$ nodes and the same number of complete subtrees. Moreover, consider a block cipher represented as $E$ and a symmetric encryption technique, impervious to IND-CPA attacks, labeled as $F$. To relay a message $M$ to $N/R$, the broadcaster emits $(i_1,...,i_m,E_{ki1}(K),... ,E_{kim}(K),F_K(M))$. Every non-revoked recipient can initially decipher the ciphertext specific to their subtree to access $K$ and subsequently unveil $M$. Under the NNL method, the ciphertext length stands at $rlogN/r$, the quantity of keys a recipient holds is $logN$, and the decryption tasks needed for a recipient remain constant at $O(1)$.

\heading{Labeled Threshold Encryption}. 
Labeled threshold encryption is a robust cryptosystem. Often simply called threshold encryption \cite{shoup1998securing}, it features a public key linked with the system while a decryption key is distributed among multiple servers. In the context of a $(t, n)$ threshold encryption scheme, which we denote as \textit{ThreshEnc}, we delineate the following algorithms. \textit{TGen}, a probabilistic key generation algorithm, ingests a security parameter $l$, the total server count $n$, and the threshold parameter $t$. It yields key components $(pk, vk, sk)$. Here, $pk$ signifies the public key, $lk$ is the verification key, and $sk = (sk_1,...,sk_n)$ is an array of private keys. \textit{TEnc}, a probabilistic encryption algorithm, requires a public key $pk$, a message $m$, and a label $l_b$, and furnishes a ciphertext $c$. \textit{ShareD}, a probabilistic decryption share generation algorithm, consumes a private key $sk_i$, a ciphertext $c$, and a label $l_b$, and produces a decryption share $\sigma$. \textit{Vrf}, a deterministic share verification algorithm, is supplied with the verification key $lk$, a ciphertext $c$, a label $l_b$, and a decryption share $\sigma$. It returns an outcome $b\in\{0, 1\}$.~\textit{Comb}, a deterministic combining algorithm, acquires the verification key $lk$, a ciphertext $c$, a label $l_b$, and a collection of $t$ decryption shares. It either outputs a message $m$ or a designated symbol $\perp$.

\heading{Trusted Execution Environment}.
 
A Trusted Execution Environment (TEE) is a secure computing environment designed to protect sensitive data and execute secure code. TEE provides secure isolation, ensuring that code and data running within the TEE are isolated from the regular operating system and its applications. This isolation protects sensitive data from unauthorized access. Gramine is a versatile tool, suitable for a myriad of confidential computing scenarios. As a library OS, it is crafted to operate applications not only on Intel® SGX but also on various other platforms. Notably, it's not restricted to the C programming language. Gramine streamlines the deployment process of applications running within enclaves. This makes it convenient for developers to launch native confidential computing applications on Intel® SGX without any alterations to the code.

\heading{Asynchronous BFT}. Asynchronous BFT protocols inherently offer robustness against timing, performance, and denial-of-service (DoS) attacks. They are, arguably, the most fitting solutions for mission-critical blockchain endeavors. Thanks to their inherent resilience, asynchronous BFT (alongside atomic broadcast) has garnered significant scholarly attention~\cite{duan2018beat}. Recent asynchronous BFT systems exhibit performance metrics on par with partially synchronous BFT protocols and have demonstrated scalability up to 100 replicas. We use the term BFT interchangeably for both Byzantine atomic broadcast and BFT without making a distinction between them.


%% file: tex/securitymodel.tex
\section{System Overview}

In this section, we first provide an overview of the system model of FACOS, and then define the threat model and introduce the design goals of our FACOS.

\subsection{FACOS System Model}

\begin{figure}[!ht]
  \centering
  \includegraphics[scale=0.045]{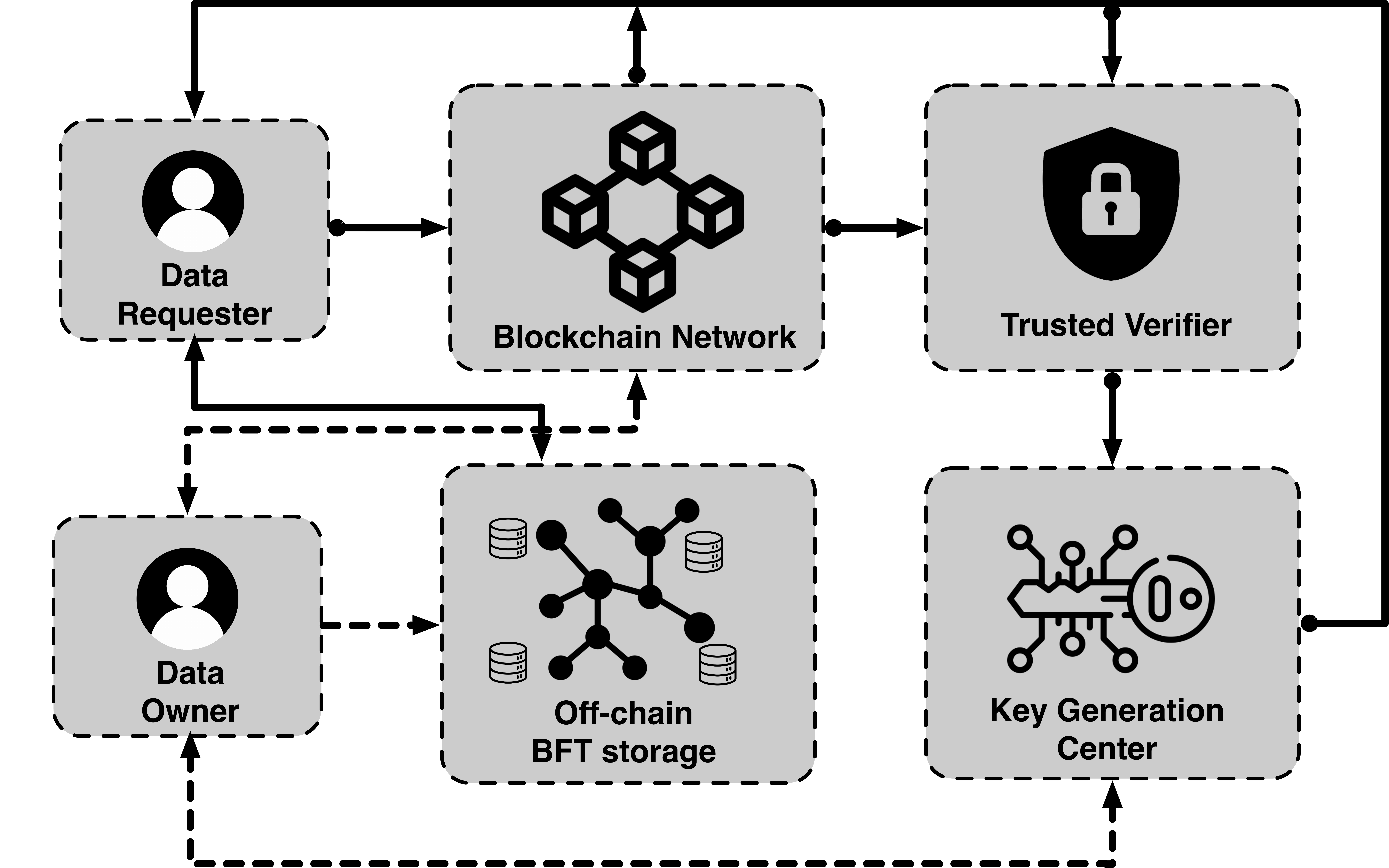}
  \caption{FACOS design overview.}
  \label{overview}
\end{figure}

In Fig. \ref{overview}, there are six entities in the system, \circlednumber{1} \textit{blockchain network}; \circlednumber{2} \textit{data owner}; \circlednumber{3} \textit{off-chain BFT storage}; \circlednumber{4} \textit{trusted verifier}; \circlednumber{5} \textit{key generation center}; \circlednumber{6} \textit{data requester}.

\circlednumber{1} \textit{Blockchain network}, in our system, we utilize the consortium blockchain network, entities that have a need to post assets or data on-chain and store them off-chain. Each consortium member is a replica licensed through an access mechanism. Each replica has a unique public/private key pair ($pk,sk$) provided by the key generation center, and the public key represents the replica’s identity. They jointly participate in and maintain the blockchain network.

\circlednumber{2} \textit{Data owner}, refers to an individual or entity who has the primary responsibility and authority over a set of data. This includes the ability to determine who can access the data, and what level of access they have. 

\circlednumber{3} \textit{Off-chain BFT storage} is the practice of storing data and transactions outside the main blockchain to alleviate its congestion and enhance scalability. It involves employing a BFT consensus algorithm to ensure data integrity and agreement among participants in a distributed network. By doing so, the system achieves greater efficiency and throughput while maintaining a high level of security and reliability. 

\circlednumber{4} \textit{Trusted verifier} via utilizing Software Guard Extensions (SGX) involves leveraging hardware-based security features provided by modern processors. The trusted verifier is utilized to establish a secure and isolated environment where critical operations can be executed, safeguarding against threats such as unauthorized access, tampering, and data breaches. In our system, the trusted verifier can help verify whether the data requester's partial policy is in the data owner's whole policy. 

\circlednumber{5} \textit{Key generation center} serves as a central entity responsible for generating and distributing cryptographic keys in a secure manner.

\circlednumber{6} \textit{Data requester} is an entity or individual that holds the authorization to request and access specific data from a source.

\subsection{Threat Model}
We describe potential threats to FACOS. We assume that the trusted verifier and key generation center are fully trusted and will not collude with any other
entity, and the data owner and data requester are honest-but-curious. That is, the data owner and data requester can honestly perform their operations and will not collude with each other or with any other malicious entity. As
data owner is responsible for providing data, and we assume that they are honest. In this paper, we focus on systems that are purely asynchronous, without making any assumptions regarding the timing of message processing or transmission delays. However, we do assume synchrony for the initial distributed key setup phase, which is a one-time event in the system's lifecycle. We make the assumption that authenticated channels are available. This is considered a minimal requirement for nodes to establish trust and authenticate with each other.

\subsection{Design goals}

Privacy emerges as a critical concern in blockchain implementation, particularly when managing sensitive data. Therefore, the FACOS should protect the message privacy, hash of message privacy, and access control policy privacy. To be specific, the message of the data owner should not be disclosed by malicious adversaries. The hash of the message of the data owner should be protected before it is sent to the blockchain. The access control policy of the data owner and the partial access control policy of the data requester should be protected.

The objective of our secure system is to attain integrity, availability, and confidentiality safeguards against compromised servers, confidentiality protection against unauthorized data requesters, and fine-grained access control. Additionally, we have established the following reliability objectives which is first defined by Chios:

\begin{itemize}
    \item \textbf{Agreement}: If any correct data owner delivers a message $m$, then every correct data requester with the same on chain $id$ delivers $m$.
    \item \textbf{Total order}: If a correct data requester has delivered $m_1,m_2, \cdot\cdot\cdot, m_s$ with an on-chain $id$ and another correct data requester has delivered $m^{'}_1,m^{'}_2, \cdot\cdot\cdot, m^{'}_s$ with the same on-chain $id$, then $m_i=m^{'}_i$ for $1\leq i\leq min(s,s^{'})$
    \item \textbf{Liveness}: If a data requester is correct and submits $m$ matching an on-chain $id$, then all correct data requesters that issued an on-chain $id$ will eventually deliver $m$. If a data requester issues an on-chain $id$, and it will receive all matching $m$.
    \item \textbf{Authentication}: If a data requester delivers a message $m$, then the
$m$ was published by some data owner.
 \item \textbf{Uniqueness}: A data requester delivers no message $m$ twice.
\end{itemize}

We consider that the off-chain storage protocol is a Byzantine fault-tolerant state machine replication (BFT) protocol, where $f$ out of $n$ replicas can fail arbitrarily (Byzantine failures).
The BFT protocols considered in this paper tolerate 
$f \leq \lfloor (n-1)/3\rfloor$ Byzantine failures, which is optimal. 

%% file: tex/system.tex
\section{FACOS: On-chain and Off-chain Solution}

This section shows the workflow and details of FACOS.

\subsection{Workflow of FACOS}

\begin{figure}[!ht]
  \centering
  \includegraphics[width=0.44\textwidth]{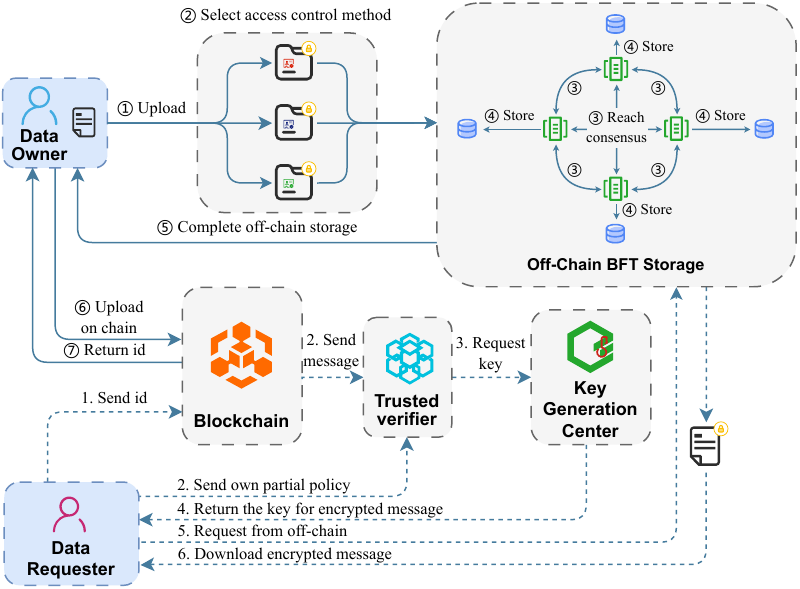}
  \caption{FACOS overflow.}
  \label{overflow}
\end{figure}

As shown in Fig. \ref{overflow}, our system includes the following steps. In the writing phase, \textcircled{\scriptsize 1} the data owner inputs a message $m$; \textcircled{\scriptsize 2} he/she selects one of the access control methods, encrypts the $m$, and finally writes the hash $h$ of message $m$ and a ciphertext $\sigma$ to off-chain BFT storage; \textcircled{\scriptsize 3} BFT storage replicas reach a consensus to this ($h$, $\sigma$); \textcircled{\scriptsize 4} BFT storage replicas store $h$ and $\sigma$ as the key-value format; \textcircled{\scriptsize 5} when the off-chain phase was done, the BFT storage replicas return ``done" to the data owner; \textcircled{\scriptsize 6} when the data owner received enough number ($f+1$) of ``done" reply from the off-chain BFT replicas, the data owner writes the access type $\mathsf{AT}$, the ciphertext of hash $h$, and the ciphertext of policy $c_p$ to the blockchain network; \textcircled{\scriptsize 7} after the on-chain phase was done, the blockchain returns $txid$ to the data owner. The data owner can share the $txid$ with some data requester at his/her willingness.


In the read phase, \textcircled{\scriptsize 1} the data requester sends the $txid$ shared by the data owner to blockchain; \textcircled{\scriptsize 2} the blockchain sends the access type  $\mathsf{AT}$ and the ciphertext of policy $c_p$ to the trusted verifier. The data requester also sends its encrypted partial attribute $c_{p_u}$ corresponding to the $txid$ to the trusted verifier; \textcircled{\scriptsize 3} After the trusted verifier compares the decrypted partial attribute with the decrypted policy, if it satisfies, the data requester can acquire the key from the key generation center in the step \textcircled{\scriptsize 4};~\textcircled{\scriptsize 5} when the data requester gets the $h$ decrypted with the key received from the key generation center, it sends $h$ to the off-chain BFT storage; \textcircled{\scriptsize 6} the data requester can download the ciphertext $\sigma$ from the off-chain BFT replicas and finally get the $m$ with using corresponding access control method.

\subsection{Details of FACOS}

\subsubsection{System setup}

\begin{table}[hbt]
	\centering
	\footnotesize
	\begin{tabular}{c|c}
		\hline
		\textbf{Symbol} & \textbf{Symbol Definition}\\
		\hline
		$\mathsf{PK_{ABE}}/\mathsf{SK_{ABE}}$/$\mathsf{MSK_{ABE}}$  & ABE key pairs/master secret key \\
		\hline
       $\mathsf{PK_{Verifier}}/\mathsf{SK_{Verifier}} $ & trusted verifier's key pairs  \\
		\hline
       $\mathsf{key_{AES}}/\mathsf{key_{SC}}$ & AES key/stream cipher key\\
        \hline
       $\mathsf{AL}$/$\mathsf{AT}$ & attribute list/access type \\
		\hline
       $\mathsf{PK_{RSA}}/\mathsf{SK_{RSA}} $ & RSA key pairs  \\
    		\hline
        $\mathsf{PK_{TE}}/\mathsf{SK_{TEi}/\mathsf{VK_{TEi}}} $ & \!\!\!\!\!\!\begin{tabular}{c}public key, secret key and verification key\\for TE  \end{tabular}\!\!\!\!\!\!\\
		\hline
        $\mathsf{P}$/$\mathsf{P_u}$ & $\mathsf{P}$ based on $\mathsf{AL}$/requester's partial attribute\\
        \hline
        $m$~/~$c_m$    &  message $m$ and ciphertext of $m$  \\
        \hline
        $h$~/~$c_h$   & hash of $m$ and ciphertext of $h$\\
        \hline
        $c_p$/$c_{p_u}$   & ciphertext of $\mathsf{P}$ and $\mathsf{P_u}$\\
        \hline
         $c_g/c_r/c_o$   & group of clients/revoked clients/XXXX\\
        \hline
	\end{tabular}
	\caption{Descryption of symbols}
	\label{tab:symbol}
\end{table}

In this phase, the off-chain BFT storage replicas, the trusted verifier, data owner, and data requester register in the key generation center. The key generation center could run $\mathsf{ABE.Setup}$, $\mathsf{AES.Setup}$, and $\mathsf{Verifier.Setup}$ algorithms and distribute corresponding keys to each entity. Table \ref{tab:symbol} lists all the symbols used.

\begin{figure}[htb]
    \begin{algorithm}[H] 
      \footnotesize
      \caption{\textbf{Owner-managed data access control protocol}}
      \label{alg1}
      \begin{algorithmic}[1]
        \STATE \textbf{INITIALIZATION}
        \begin{ALC@g}
          \STATE $buf$ $\leftarrow \emptyset$ \hfill \{message buffer\}
          \STATE $m$ $\leftarrow buf$ \hfill \{select first batch of message from buffer\}
          \STATE $\mathsf{AL}$, $\mathsf{AT}$ $\leftarrow$ \{1:$\mathsf{BE}$, 2:$\mathsf{AE}$, 3:$\mathsf{TE}$\} and $\mathsf{P}$ \hfill \{attribute list, access type and policy\}
          \STATE let $\mathsf{PK_{Verifier}}$ be the public key received from the setup step by $\mathsf{Verifier.Setup}$
        \end{ALC@g}
        \STATE \textbf{UPON} receiving ($m$, $\mathsf{AT}$, $\mathsf{AL}$) \textbf{DO}
        \begin{ALC@g}
          \STATE $\mathsf{P}$ $\leftarrow$ $\mathsf{AL}$ \hfill \{generate $\mathsf{P}$ based on $\mathsf{AL}$\}
          \STATE pass ($m$, $\mathsf{AT}$, $\mathsf{P}$) into FACOS WRITE Engine \hfill \{{\textcolor[rgb]{1,0,0}{$\triangleright$} see Alg.~\ref{alg2}}\}\,\,
        \end{ALC@g}
        \STATE \textbf{UPON} receiving ($c_m$, $x$, $c_h$) from FACOS Write Engine \textbf{DO}
        \begin{ALC@g}
          \STATE generate $h$ := $\mathsf{Hash}$($m$)
          \STATE pack $\sigma$ := ($c_m$, $x$)
          \STATE send ($h$, $\sigma$) to off-chain BFT storage replica  randomly \hfill \{{\textcolor[rgb]{1,0,0}{$\triangleright$} see Alg.~\ref{alg3}}\}
        \end{ALC@g}
        \STATE \textbf{UPON} delivery of $\mathsf{True}$ from ~$f+1$~ BFT storage~ replicas \textbf{DO}
        \begin{ALC@g}
          \STATE encrypt $c_{p}$ := $\mathsf{Verifier.Enc}$($\mathsf{PK_{Verifier}}$, $\mathsf{P}$)
          \STATE pack $\alpha$ := ($write$, $\mathsf{AT}$, $c_h$, $c_{p}$)
          \STATE send $\alpha$ to blockchain network
        \end{ALC@g}
        \STATE \textbf{UPON} delivery of $txid$ from blockchain network \textbf{DO}
        \begin{ALC@g}
          \STATE share message $txid$ with data requester or store it locally 
        \end{ALC@g}
      \end{algorithmic}
    \end{algorithm}
    \caption{Outlined the execution process from the data owner's perspective.}
\end{figure}

\subsubsection{Write phase}
This part includes three building blocks: \circlednumber{1} \textit{Owner-managed data access control protocol}; \circlednumber{2} \textit{Off-chain consensus and storage protocol}; \circlednumber{3} \textit{On-chain write protocol}.

\circlednumber{1} \textit{Owner-managed data access control protocol}

The algorithm described in Alg. \ref{alg1} focuses on the perspective of data owners to depict the entire process of data off-chain and on-chain write phases. The 1-5 lines represent the system initialization stage. From lines 6 to 8, when the data owner inputs ($m$, $\mathsf{AT}$, $\mathsf{AL}$), he/she passes it into the core part of FACOS, which we refer to as the FACOS WRITE Engine. The data owner can choose corresponding fine-grained access control methods based on their different requirements to set access control for data information. Alg. \ref{alg2} displays the core algorithm of the FACOS WRITE Engine. The initialization stage focuses on generating keys for three different building blocks. From lines 6 to 12, when receiving the message ($m$, $\mathsf{AT}$, $\mathsf{P}$) from Alg. \ref{alg1}, the FACOS WRITE Engine selects different access control methods based on the access control mode chosen by $\mathsf{AT}$. The core code of FACOS with ABE module, shown in lines 14-18, utilizes a hybrid encryption approach. It employs symmetric encryption to generate $c_m$ for the message $m$ and employs attribute-based encryption to encrypt the symmetric encryption key $\mathsf{key_{AES}}$. Similarly, the Hash value $h$ of the message $m$ is encrypted using attribute-based encryption. This approach ensures message confidentiality, even if an adversary obtains the value corresponding to $txid$ on the blockchain because the critical values of the value, namely the key values in the off-chain, are also ciphertexts. Finally, ($c_m$, $x$, $c_h$) are returned to Alg. 1. For FACOS with BE module, the data owner inputs a group of clients, and its size is $n$ at line 24. The revoked clients $\{cr_i,\ldots, cr_j\}$ are shown in line 25. These are $n$ clients that can generate a full binary tree and be as leaves at line 27. The generation process of a full binary tree

\begin{figure}[htb]
    \begin{algorithm}[H] 
      \footnotesize
      \caption{\textbf{FACOS WRITE Engine}}
      \label{alg2}
      \begin{algorithmic}[1]
        \STATE \textbf{INITIALIZATION}
        \begin{ALC@g}
          \STATE Let $\mathsf{PK_{ABE}}$ be the public key received from $\mathsf{ABE.Setup}$, and let $\mathsf{MSK_{ABE}}$ be the master secret key
          \STATE Let $\mathsf{key_{AES}}$ be the key for AES algorithm
          \STATE Let $\mathsf{key_{SC}}$ be the key for stream cipher algorithm
          \STATE Let $\mathsf{PK_{TE_i}}$ be the public key received from $\mathsf{TE_{Setup}}$
        \end{ALC@g}
        \STATE \textbf{UPON} receiving ($m$, $\mathsf{AT}$, $\mathsf{P}$) \textbf{DO}
        \begin{ALC@g}
          \STATE \textbf{switch}($\mathsf{AT}$) \textbf{do}
          \begin{ALC@g}
            \STATE case ABE: ($c_m$, $x$, $c_h$):=$\mathsf{AccessControlABE}$($m$, $\mathsf{AL}$, $\mathsf{P}$)
            \STATE case BE:~~~($c_m$, $x$, $c_h$):=$\mathsf{AccessControlBE}$($m$, $\mathsf{P}$)
            \STATE case TE:~~~($c_m$, $x$, $c_h$):=$\mathsf{AccessControlTE}$($m$, $\mathsf{P}$)
            \STATE case otherwise: $\mathsf{Quit()}$
          \end{ALC@g}
          \STATE \textbf{end switch}
          \STATE \textbf{return} ($c_m$, $x$, $c_h$)
        \end{ALC@g}
        \hdashrule[0.6ex]{\linewidth}{0.6pt}{1.5mm}%
        \STATE \textbf{PROCEDURE} $ \mathsf{AccessControlABE}$($m$, $h$, $\mathsf{AL}$, $\mathsf{P}$)\hfill \{{\textcolor[rgb]{1,0,0}{$\triangleright$} FACOS with ABE module}\}
        \begin{ALC@g}
          \STATE encrypt $c_m:=\mathsf{AES.Enc}$($\mathsf{key_{AES}}$, $m$)
          \STATE encrypt $x:=\mathsf{ABE.Enc}$($\mathsf{PK_{ABE}}$, $\mathsf{key_{AES}},$ $\mathsf{P}$)
          \STATE encrypt $c_h:=\mathsf{ABE.Enc}$($\mathsf{PK_{ABE}}$, $h$, $\mathsf{P}$)
          \STATE \textbf{output} ($c_m$, $x$, $c_h$)
        \end{ALC@g}
        \STATE \textbf{PROCEDURE} $\mathsf{AccessControlBE}$($m$, $h$, $\mathsf{P}$) \hfill \{{\textcolor[rgb]{1,0,0}{$\triangleright$} FACOS with BE module}\}
        \begin{ALC@g}
          \STATE ($c_m$, $x$) := $\mathsf{EncryptBE}$($m$, $\mathsf{P}$)
          \STATE $c_h$ := $\mathsf{RSA.Enc}$($\mathsf{PK_{RSA}}$, $h$)
          \STATE \textbf{output} ($c_m,x$, $c_h$)
        \end{ALC@g}
        \STATE \textbf{PROCEDURE} $\mathsf{EncryptBE}$($m$, $\mathsf{P}$)
        \begin{ALC@g}
          \STATE Let $c_g=\{cg_1,cg_{2},\ldots, cg_{n}\}$ be a group of clients \hfill \{$n$ clients\}\,\,
          \STATE Let $c_r=\{cr_i,\ldots, cr_j\}$ be revoked clients
          \STATE Let $c_o = c_g/c_r= \{co_l,\ldots,co_{i-1},co_{j+1},\ldots,co_r\}$\\ be the clients who can receive $\text{message}$
          \STATE a group of clients \{$cg_1,cg_{2},\ldots,cg_n$\} generate a full\\ binary tree and be as leaves 
          \STATE $h=\mathsf{GenerationFullBinaryTree}$($c_g$, $c_r$, $c_o$, $n$) \hfill \{$h$ points to the root of a full binary tree\}
          \STATE $c_m$ := $\mathsf{StreamCipher.Enc}$($\mathsf{key_{SC}}$, $m$)
          \STATE $\mathsf{P}$ = $\{l,\ldots,i-1,j+1,\ldots,r\}$
          \STATE $\mathsf{K}$ := $\mathsf{P}$
          \STATE \textbf{for} $i \leftarrow {n-2}\ {to}\ 0:$
          \begin{ALC@g}
            \STATE \textbf{if} $2i+1$\ in $\mathsf{K}$\ and $2i+2$ in $\mathsf{K}$:
            \begin{ALC@g}
              \STATE delete $2i+1$ and $2i+2$ in $\mathsf{K}$
              \STATE add $i$ in $\mathsf{K}$
            \end{ALC@g}
          \end{ALC@g}
          \STATE \textbf{for} $i \in$ $\mathsf{K}$:
          \begin{ALC@g}
            \STATE encrypt $c_i$ := $\mathsf{AES.Enc}$($k_i$, $\mathsf{key_{SC}}$)
          \end{ALC@g}
          \STATE $x$ := [$c_i$($i$ $\in$ $\mathsf{K}$)]
          \STATE \textbf{output} ($c_m$, $x$)
        \end{ALC@g}
        \STATE \textbf{PROCEDURE} $ \mathsf{GenerationFullBinaryTree}$($c_g$, $c_r$, $c_o$, $n$)
        \begin{ALC@g}
          \STATE generate a full binary tree with ($2 \times n-1$) nodes
          \STATE the root is numbered with 0
          \STATE the key of the $i$-th node is $k_i$ ($i$ $\in$ [$0,2\times n-2$])
          \STATE a leaf node has a key list $\{k_i,k_{\lfloor(i+1)/2\rfloor},\ldots, k_0\}$  \hfill \{a leaf node owns its ancestors' keys\}
          \STATE $n$ leaves represent clients
          \STATE \textbf{output} $h$
        \end{ALC@g}
        \STATE \textbf{PROCEDURE} $\mathsf{AccessControlTE}$($m$, $h$, $\mathsf{P}$)  \hfill \{{\textcolor[rgb]{1,0,0}{$\triangleright$} FACOS with TE module}\}
        \begin{ALC@g}
          \STATE encrypt $c_m$ := $\mathsf{AES.Enc}$($\mathsf{key_{AES}}$, $m$)
          \STATE encrypt $x$ := $\mathsf{TE.Enc}$($\mathsf{PK_{TE}}$, $\mathsf{key_{AES}}$, $\mathsf{P}$)
          \STATE $c_h$ := $\mathsf{RSA.Enc}$($\mathsf{PK_{RSA}}$, $h$)
          \STATE \textbf{output} ($c_m$, $x$, $c_h$)
        \end{ALC@g}
      \end{algorithmic}
    \end{algorithm}
    \caption{Iustration of the FACOS WRITE Engine algorithm for data owner message processing.}
\end{figure}

\noindent is shown in lines 40 to 46. Each leaf node can obtain its key $k_i$ and its ancestors' keys at line 44. At line 37, the data owner encrypts the message $m$ with the stream cipher algorithm by $\mathsf{key_{SC}}$ and then encrypts the $\mathsf{key_{SC}}$ by the $k_i$ which in the final $K$. Generating the $K$ is from line 30 to line 35. If $i$'s two children nodes $2i+1$ and $2i+2$ in $K$, it deletes these two children nodes and adds their parent node $i$ in $K$. With the above recursion, the final list $K$ will be generated at line 35. This $K$ includes the data requester's attribute which can encrypt the $\mathsf{key_{SC}}$. From line 47 to line 51, the FACOS with TE module adopts a threshold encryption scheme, with the same process as FACOS with ABE module.

When the data owner received the return values ($c_m$, $x$, $c_h$) from the FACOS WRITE Engine, the data owner first completes off-chain storage, which occurs during lines 9-12. The data owner generates the corresponding hash value $h$ for $m$, then packs and returns the values ($c_m$, $x$), storing ($h$, $\sigma$) off-chain.

\circlednumber{2} \textit{Off-chain consensus and storage protocol}

We utilize the asynchronous BFT protocol as the off-chain storage protocol. In FACOS, this modular can be served as a black box. When the message ($h$, $\sigma$) inputs into this black box, it can be finally stored at least $2f+1$ correct replicas. As depicted in Alg. \ref{alg3}, we take the HALE~\cite{zhang2022achieve} protocol which is based on BKR \cite{ben1994asynchronous} framework for an example. The BKR framework consists of two phases: the reliable broadcast (RBC) phase and the asynchronous binary agreement (ABA) phase. The framework is leaderless and all replicas propose transactions in parallel. The protocol proceeds in epochs, where in each epoch replicas reach an agreement on the order of several batches of transactions, proposed concurrently by all replicas.

\begin{figure}[htb]
    \begin{algorithm}[H] 
      \footnotesize
      \caption{\textbf{BFT Storage Protocol}}
      \label{alg3}
      \begin{algorithmic}[1]
        \STATE \textbf{INITIALIZATION}
        \begin{ALC@g}
          \STATE $buf$ $\leftarrow \emptyset$ receiving message from data owner \hfill \{transaction buffer\}
          \STATE $\mathsf{B}$ \hfill \{batch size\}
          \STATE $s \leftarrow 0$ \hfill \{epoch number\}
          \STATE $i$ \hfill \{replica id\}
          \STATE \{RBC$_j$\}$_n$ \hfill \{$n$ RBC instances where $j$ is the sender of RBC$_j$\}
          \STATE \{ABA$_j$\}$_n$ \hfill \{$n$ ABA instances\}\,\,
        \end{ALC@g}
        \STATE \textbf{EPOCH $\emph{s}$}
        \begin{ALC@g}
          \STATE Let $value$ be the first $\lceil $$\mathsf{B}$$/n \rceil$ transactions in $buf$
          \STATE Input $value$ to RBC$_i$
          \STATE \textbf{UPON} delivery of $value_j$ from RBC$_j$
          \begin{ALC@g}
            \STATE \textbf{if} ABA$_j$ has not yet been provided input, input $1$ to ABA$_j$
          \end{ALC@g}
          \STATE \textbf{UPON} delivery of $1$ from ABA$_j$ and $value_j$ from RBC$_j$
          \begin{ALC@g}
            \STATE $output \leftarrow output \cup value_j$
          \end{ALC@g}
          \STATE \textbf{UPON} delivery of $1$ from at least $n-f$ ABA instances
          \begin{ALC@g}
            \STATE \textbf{for} each ABA$_j$ instance that has not been provided input
            \begin{ALC@g}
              \STATE input $0$ to ABA$_j$
            \end{ALC@g}
          \end{ALC@g}
          \STATE \textbf{UPON} termination of all the $n$ ABA instances
          \begin{ALC@g}
            \STATE deliver $output$
            \STATE store ($h$, $c$) in its local database
            \STATE return $\mathsf{True}$ to data owner
          \end{ALC@g}
          \STATE $s \leftarrow s+1$
        \end{ALC@g}
      \end{algorithmic}
    \end{algorithm}
    \caption{Off-chain BFT Storage for replica $p_i$.}
\end{figure}

Each epoch replica $p_i$ ($i \in [0..n-1]$) proposes a batch of transactions $tx_i$ at line 9. Replica $p_i$ uses RBC to broadcast the proposals at lines 10. After the RBC phase, $n$ parallel ABA instances will be triggered where the $i$-th ABA instance is used for replicas to agree on whether the $i$-th proposal has been delivered in the RBC phase. Specifically, each replica will vote for 1 if the corresponding proposal has been delivered and 0 otherwise at lines 15-17. In other words, $n$ parallel RBC instances and $n$ parallel ABA instances are included in each epoch. If an ABA instance terminates with 1, replicas will add the corresponding proposal (transactions) to its delivery queue. After all $n$ ABA instances have been terminated, non-overlapped transactions in the delivery queue of each replica will be delivered according to a deterministic order.

Finally, each correct replica stores the ($h$, $\sigma$) as a key-value in its local database, returns $\mathsf{True}$ message to the data owner, and starts the next epoch. The off-chain phase is done.

\circlednumber{3} \textit{On-chain write protocol}

We treat blockchain as a black box, and we illustrate the process of the on-chain phase in Alg. \ref{alg1}. In Alg. \ref{alg1}, when the data owner received $f+1$ ``$\mathsf{True}$" messages from BFT storage replicas, it means that the off-chain phase was done for the data owner, and the data owner can start on-chain protocol at line 13. The data owner encrypts the policy $\mathsf{P}$ with the trusted verifier's public key $\mathsf{PK_{Verifier}}$ at line 14, packs the $write$ command, the access type $\mathsf{AT}$, the hash $h$' ciphertext $c_h$, and ciphertext of policy $c_p$ into $\alpha$ at line 15, and finally sends this $\alpha$ to the blockchain at line 16. 

\begin{figure}[htb]
    \begin{algorithm}[H] 
      \footnotesize
      \caption{\textbf{Requester-driven Data Retrieval Protocol}}
      \label{alg4}
      \begin{algorithmic}[1]
        \STATE \textbf{INITIALIZATION}
        \begin{ALC@g}
          \STATE $txid$  \hfill \{message $txid$ shared by data owner\}
          \STATE $\mathsf{P_u}$ \hfill \{personal attribute\}\,\,
        \end{ALC@g}
        \STATE \textbf{UPON} receiving $txid$ \textbf{DO}
        \begin{ALC@g}
          \STATE $c_{p_u}$ := $\mathsf{Verifier.Enc}$($\mathsf{PK_{Verifier}}$, $\mathsf{P_u}$)
          \STATE send $txid$ to blockchain network
          \STATE send $c_{p_u}$ to the trusted verifier \hfill \{{\textcolor[rgb]{1,0,0}{$\triangleright$} see Alg.~\ref{alg6}}\}
          \STATE fetch $c_{h}$ from blockchain network
        \end{ALC@g}
        \STATE \textbf{UPON} receiving $\mathsf{key}$ from key generation center
        \begin{ALC@g}
          \STATE decrypt $h$ := $\mathsf{DecryptGetHash}$($\mathsf{AT}$, $\mathsf{key}$)
          \STATE send $h$ to BFT storage replicas
        \end{ALC@g}
        \STATE \textbf{UPON} delivery of $\sigma$ from $f+1$ BFT storage replicas
        \begin{ALC@g}
          \STATE unpack ($c_m$, $x$) := $\sigma$
          \STATE pass ($c_m$, $x$) to FACOS READ Engine\hfill \{{\textcolor[rgb]{1,0,0}{$\triangleright$} see Alg.~\ref{alg5}}\}\,\,
          \STATE deliver $m$
        \end{ALC@g}
        \hdashrule[0.6ex]{\linewidth}{0.6pt}{1.5mm}%
        \STATE \textbf{PROCEDURE} $\mathsf{DecryptGetHash}$($\mathsf{AT}$, $\mathsf{key}$)
        \begin{ALC@g}
          \STATE \textbf{switch}($\mathsf{AT}$) \textbf{do}
          \begin{ALC@g}
            \STATE case ABE:
            \begin{ALC@g}
              \STATE $\mathsf{SK_{ABE}}$ := $\mathsf{key}$
              \STATE decrypt $h$ := $\mathsf{ABE.Dec}$($\mathsf{PK_{ABE}}$, $c_h$, $\mathsf{SK_{ABE}}$)
            \end{ALC@g}
            \STATE case BE:
            \begin{ALC@g}
              \STATE $\mathsf{SK_{RSA}}$ := $\mathsf{key}$
              \STATE decrypt $h$ := $\mathsf{RSA.Dec}$($\mathsf{SK_{RSA}}$, $c_h$)
            \end{ALC@g}
            \STATE case TE:
            \begin{ALC@g}
              \STATE $\mathsf{SK_{RSA}}$ := $\mathsf{key}$
              \STATE decrypt $h$ := $\mathsf{RSA.Dec}$($\mathsf{SK_{RSA}}$, $c_h$)
            \end{ALC@g}
            \STATE case otherwise: $\mathsf{Quit()}$
          \end{ALC@g}
          \STATE \textbf{end switch}
          \STATE \textbf{return} ($h$)
        \end{ALC@g}
      \end{algorithmic}
    \end{algorithm}
    \caption{Outlined the execution process from the data requester's perspective.}
\end{figure}

We use the permissioned blockchain to proceed the on-chain phase. We illustrate the on-chain phase for simplicity.

Upon receiving the data owner's message $proposed$, the blockchain unpacks it and stores it within a new transaction. This transaction includes the message ($write$, $\mathsf{AT}$, $c_h$, $c_p$) along with relevant metadata, such as a timestamp. The new transaction is broadcast to nodes in the consortium blockchain network. Nodes validate the transaction's legitimacy and include it in a new block. Nodes in the consortium blockchain reach consensus through a consensus algorithm to validate the transaction and create a new block. Once consensus is achieved, the new block is added to the blockchain. The blockchain network generates a unique message $txid$~(typically a hash value) to identify the message ($write$, $\mathsf{AT}$, $c_h$, $c_p$) on the blockchain. The blockchain application or interface sends the message $id$ back to the data owner, confirming that the message ($write$, $\mathsf{AT}$, $c_h$, $c_p$) has been successfully stored on the blockchain. The data owner can now use this $txid$ to retrieve the message or verify its existence, and he/she can share the $txid$ with some data requester or store it locally.

\subsubsection{Read phase}

This part consists of four building blocks: \circlednumber{1} \textit{Requester-driven data retrieval protocol}; \circlednumber{2} \textit{On-chain read protocol}; \circlednumber{3} \textit{Trusted verifier verification protocol}; \circlednumber{4} \textit{Off-chain read phase}.

\circlednumber{1} \textit{Requester-driven data retrieval protocol}

In Alg. \ref{alg4}, upon receiving the $txid$ shared by the data owner, the data requester can access the message $m$, provided they have the appropriate permissions, as determined by completing the four steps of the read phase. First, a data requester encrypts its partial access attribute with the trusted verifier's public key $\mathsf{PK_{Verifier}}$ at line 5, and sends $txid$ to blockchain and its encrypted attribute to the trusted verifier from line 7 to line 8, respectively.

\begin{figure}[htb]
    \begin{algorithm}[H] 
      \footnotesize
      \caption{\textbf{FACOS READ Engine}}
      \label{alg5}
      \begin{algorithmic}[1]
        \STATE \textbf{INITIALIZATION}
        \begin{ALC@g}
          \STATE \textbf{UPON} receiving ($c_m$, $x$, $\mathsf{AT}$) \textbf{DO}
          \begin{ALC@g}
            \STATE \textbf{switch}($\mathsf{AT}$) \textbf{do}
            \begin{ALC@g}
              \STATE case ABE:  $m:= \mathsf{AccessControlABE}$($c_m$, $x$)
              \STATE case BE: $m:=\mathsf{AccessControlBE}$($c_m$, $x$)
              \STATE case TE $m:=\mathsf{AccessControlTE}$($c_m$, $x$)
              \STATE case otherwise: $\mathsf{Quit()}$
            \end{ALC@g}
            \STATE \textbf{end switch}
            \STATE \textbf{return} $m$
          \end{ALC@g}
        \end{ALC@g}
        \hdashrule[0.6ex]{\linewidth}{0.6pt}{1.5mm}%
        \STATE \textbf{PROCEDURE} $ \mathsf{AccessControlABE}$($c_m$, $x$) \hfill \{{\textcolor[rgb]{1,0,0}{$\triangleright$} FACOS with ABE module}\}
        \begin{ALC@g}
          \STATE decrypt $\mathsf{key_{AES}}$ := $\mathsf{ABE.Dec}$($\mathsf{PK_{ABE}}$, $c_h$, $\mathsf{SK_{ABE}})$
          \STATE decrypt $m$ := $\mathsf{AES.Dec}$($\mathsf{key_{AES}}$, $x$)
          \STATE \textbf{output} ($m$)
        \end{ALC@g}
        \STATE \textbf{PROCEDURE} $\mathsf{AccessControlBE}$($c_m$, $x$) \hfill \{{\textcolor[rgb]{1,0,0}{$\triangleright$} FACOS with BE module}\}
        \begin{ALC@g}
          \STATE find $c_i$ in right location of $c$ based on $p_u$
          \STATE decrypt $\mathsf{key_{SC}}$ := $\mathsf{AES.Enc}$($k_i$, $c_i$)
          \STATE $m$ := $\mathsf{StreamCipher.Enc}$($\mathsf{key_{SC}}$, $x$)
          \STATE \textbf{output} ($m$)
        \end{ALC@g}
        \STATE \textbf{PROCEDURE} $\mathsf{AccessControlTE}$($c_{m}$, $x$) \hfill \{{\textcolor[rgb]{1,0,0}{$\triangleright$} FACOS with TE module}\}
        \begin{ALC@g}
          \STATE $count$ := 0
          \STATE unpack ($c_m$, $share_i$) := ($c_m$, $x$)
          \STATE \textbf{if} $\mathsf{TE.Ver}$($\mathsf{VK_{TE_i}}$, $share_i$) == $\mathsf{True}$
          \begin{ALC@g}
            \STATE $count$ := $count+1$
            \STATE \textbf{if} $count \geq f+1$
            \begin{ALC@g}
              \STATE combine $\mathsf{key_{AES}}$ := $\mathsf{TE.Com}$($\mathsf{PK_{TE}}$, $share_i$)
            \end{ALC@g}
          \end{ALC@g}
          \STATE decrypt $m$ := $\mathsf{AES.Dec}$($\mathsf{key_{AES}}$, $c_m$)
          \STATE \textbf{output} ($m$)
        \end{ALC@g}
      \end{algorithmic}
    \end{algorithm}
    \caption{Illustration of the FACOS READ Engine algorithm for data owner message processing.}
\end{figure}

From line 9 to line 10, when the data requester receives the key $\mathsf{key}$ from the key generation center, from line 16 to line 29, the data requester decrypts the received ciphertext $c_h$ from the blockchain based on the selected access control scheme, ultimately obtaining the hash value $h$. It is noteworthy that, due to the complexity and limitations of the decryption algorithms, only FACOS with ABE module employs an ABE algorithm, while the other two methods utilize RSA encryption schemes.

From line 12 to line 13, when the data requester receives $f+1$ identical signatures $\sigma$, the data requester unpacks $\sigma$ to obtain ($c_m$, $x$). Finally, ($c_m$, $x$) is input into Alg. \ref{alg5}, FACOS READ Engine, to obtain the final message $m$. In Alg. \ref{alg5}, the process of the FACOS READ Engine is described. Here, the emphasis is on detailing the FACOS with TE module protocol. From line 19 to line 27, due to the unique nature of the threshold encryption read phase, referencing the off-chain read phase, each off-chain BFT replica returns a share of the ciphertext. Upon receiving the shares from off-chain BFT replica at line 21, the algorithm first verifies the correctness of the shares. Upon successful verification of the shares, at line 22, the counter $count$ is incremented. When the $count$ exceeds $f+1$, the data requester utilizes the TE Combine algorithm to finally reconstruct the key $\mathsf{key_{AES}}$. Finally, the message $m$ is decrypted at line 26.

\circlednumber{2} \textit{On-chain read protocol}

Blockchain received $txid$ from the data requester, afterward, the blockchain sends the ($\mathsf{AT}$, $c_h$) to the data requester and ($\mathsf{AT}$, $c_p$) to the trusted verifier. We'll not describe the detailed steps.

\begin{figure}[htb]
    \begin{algorithm}[H] 
      \footnotesize
      \caption{\textbf{Trusted Verifier Verification Protocol}}
      \label{alg6}
      \begin{algorithmic}[1]
        \STATE \textbf{INITIALIZATION}
        \begin{ALC@g}
          \STATE $res \leftarrow$ false \hfill \{response to key generation center\}
          \STATE $c_{p}$
          \STATE $\mathsf{P_u}$
          \STATE $\mathsf{AT}$ $\leftarrow$ \{1:$\mathsf{BE}$, 2:$\mathsf{AE}$, 3:$\mathsf{TE}$\} \hfill \{trusted\\ verifier's access control type\}
          \STATE Let $\mathsf{SK_{Verifier}}$ be the secret key from $\mathsf{Verifier.Setup}$
        \end{ALC@g}
        \STATE \textbf{UPON} receiving of ($\mathsf{AT}$$_b$, $c_{p}$) from blockchain and ($\mathsf{AT}$$_d$, $c_{p_u}$) from data requester   \hfill \{$\mathsf{AT}$$_b$ and $\mathsf{AT}$$_d$ mean the access control method received from blockchain and data requester, respectively.\}
        \begin{ALC@g}
          \STATE decrypt $\mathsf{P}$ = $\mathsf{Verifier.Dec}$($\mathsf{SK_{Verifier}}$, $c_{p}$)
          \STATE decrypt $\mathsf{P_{u}}$ = $\mathsf{Verifier.Dec}$($\mathsf{SK_{Verifier}}$, $c_{p_u}$)
          \STATE $\mathsf{if}$ ($\mathsf{AT}$ == $\mathsf{AT}$$_b$) and ($\mathsf{AT}$ == $\mathsf{AT}$$_d$)
          \begin{ALC@g}
            \STATE $\mathsf{if}$ $\mathsf{P_u}$ satisfies $\mathsf{P}$'s syntax
            \begin{ALC@g}
              \STATE $res \leftarrow$ $\mathsf{true}$
            \end{ALC@g}
          \end{ALC@g}
         \STATE send $res$ to key generation center
        \end{ALC@g}
      \end{algorithmic}
    \end{algorithm}
    \caption{Trusted verifier verification phase.}
\end{figure}

\circlednumber{3} \textit{Trusted verifier verification protocol}

In Alg. \ref{alg6}, when the trusted verifier received the $c_p$ from blockchain and $c_{p_u}$ from the data requester at line 7, the trusted verifier utilizes its private key to decrypt the $c_p$ and $c_{p_u}$ to get the $\mathsf{P}$ and $\mathsf{P_u}$ at line 8 to 9. If they are in the same access type and the $\mathsf{P_u}$ satisfies the $\mathsf{P}$'s syntax, the trusted verifier returns the ``$\mathsf{True}$" to the key generation center.

\circlednumber{4} \textit{Off-chain read phase}

The off-chain BFT replication read phase is relatively straightforward. The data requester sends $h$ to each replica, and then each replica sends the corresponding $\sigma$ to the data requester. The FACOS with TE module differs slightly from other protocols. Upon receiving a read command, each replica extracts $\sigma$ and unpacks it to obtain ($c_m$, $x$), where the replica utilizes its private key $\mathsf{SK_{TEi}}$ to apply the $\mathsf{TE.Dec}$ algorithm to $x$, resulting in $share_i$. Finally, the data owner repackages ($c_m$, $share_i$) and generates a new $\sigma$ to send to the data requester.

%% file: tex/security.tex
\section{Theoretical Analysis}
\subsection{ Privacy analysis}
\par In this subsection, we prove that the FACOS can preserve the message privacy, hash of message privacy, and access control policy privacy.

\textbf{1) Message privacy}

\hangafter 1 \hangindent 1em \noindent {\textbf{Theorem 1. }} Since the data owner encrypts message $m$ and uploads the encrypted message to off-chain BFT storage, the FACOS will not disclose the message $m$ of the data owner, that is, any PPT adversary $\mathcal{A}$ learns nothing about the message $m$.

\textbf{Proof.} If a PPT adversary $\mathcal{A}$ desires to learn the message $m$ from the ciphertext $c_m$ := $\mathsf{AES.Enc}$($\mathsf{key_{AES}}$, $m$), then $\mathcal{A}$ needs to decrypt ciphertext $c_m$. In this case, $\mathcal{A}$ cannot crack the ciphertext $c_m$ unless $\mathcal{A}$ can decrypt ciphertext $x$ := $\mathsf{ABE.Enc}$($\mathsf{PK_{ABE}}$, $\mathsf{key_{AES}}$, $\mathsf{P}$) to obtain the key $\mathsf{key_{AES}}$. However, according to \cite{agrawal2017fame}, the CPA-ABE algorithm used in this paper is fully secure under the DLIN assumption on asymmetric pairing groups in the random oracle model. Concretely, for any PPT adversary $\mathcal{A}$ making $Q$ key queries in the IND-CPA security game, there exists a PPT adversary $\mathcal{B}$ such that $$ Adv_{CPA-ABE}^{\mathcal{A}}(\lambda) \leq (8Q+2) Adv_{DLIN}^{\mathcal{B}}(\lambda)+(16Q+6)/p$$
where $Q$ is the total number of key queries $\mathcal{A}$ makes, $p = \Theta(\lambda)$ is the order of the pairing group. Therefore, any PPT adversary $\mathcal{A}$ learns nothing about the message $m$.

\textbf{2) Hash of message privacy}

\hangafter 1 \hangindent 1em \noindent {\textbf{Theorem 2. }} Since the data owner sends the ciphertext of hash to the blockchain, the FACOS will not disclose the hash of the message of the data owner, that is, any PPT adversary $\mathcal{A}$ learns nothing about the hash of the message.

\textbf{Proof.} If a PPT adversary $\mathcal{A}$ desires to learn the hash of message from the ciphertext $c_h$ := $\mathsf{ABE.Enc}$($\mathsf{PK_{ABE}}$, $h$, $\mathsf{P}$), then $\mathcal{A}$ needs to decrypt ciphertext $c_h$. In this case, for the same reason mentioned above, CPA-ABE is fully secure under the DLIN assumption on asymmetric pairing groups in the random oracle model. Therefore, any PPT adversary $\mathcal{A}$ learns nothing about the hash of the message.

\textbf{3) Access control policy privacy}

\hangafter 1 \hangindent 1em \noindent {\textbf{Theorem 3. }} Since the data owner encrypts the access control policy with the trusted verifier's public key $\mathsf{PK_{Verifier}}$ and data requester encrypts its partial access control policy with the trusted verifier's public key $\mathsf{PK_{Verifier}}$, the FACOS will not disclose the access control policy of the data owner and partial access control policy of the data requester, that is, any PPT adversary $\mathcal{A}$ can learn nothing about the access control policy.

\textbf{Proof.} If a PPT adversary $\mathcal{A}$ desires to learn the access control policy of the data owner from the ciphertext $c_p$ := $\mathsf{Verifier.Enc}$($\mathsf{PK_{Verifier}}$, $\mathsf{P}$) and partial access control policy of the data requester from the ciphertext $c_{p_u}$:= $\mathsf{Verifier.Enc}$($\mathsf{PK_{Verifier}}$, $c_{p_u}$), then $\mathcal{A}$ needs to decrypt ciphertexts $c_p$ and $c_{p_u}$. In this case, $\mathcal{A}$ cannot crack the ciphertexts $c_p$ and $c_{p_u}$ unless $\mathcal{A}$ is able to obtain the key $\mathsf{SK_{Verifier}}$. However, it is difficult for $\mathcal{A}$ to obtain the key $\mathsf{SK_{Verifier}}$ because the integer factorization problem is hard. Therefore, any PPT adversary $\mathcal{A}$ learns nothing about the access control policy.

\subsection{Security analysis}
\textbf{Proof.} We now show the correctness of FACOS.

\begin{table*}[!ht]
	\centering
	\scriptsize
	\begin{tabular}{|c|c|c|c|c|c|c|c|}
		\hline
		Protocol  &\!\!\!\!\!\!\!\! \begin{tabular}{c}Fine-grained\\ access control\end{tabular}\!\!\!\!\!\!\!\! & \!\!\!\!\begin{tabular}{c}Practical\\ and scalable\end{tabular}\!\!\!\! &\!\!\!\!\!\!\!\!\begin{tabular}{c}Formal\\ proof\end{tabular}\!\!\!\!\!\!\!\! 
       &\!\!\!\!\!\!\!\!\begin{tabular}{c}Byzantine\\ off-chain storage\end{tabular}\!\!\!\!\!\!\!\!
       &\!\!\!\!\!\!\!\!\begin{tabular}{c}Data\\ security\end{tabular}\!\!\!\!\!\!\!\!
       &\!\!\!\!\!\!\!\!\begin{tabular}{c}Data privacy \\ protection\end{tabular}\!\!\!\!\!\!\!\!
       &\!\!\!\!\!\!\!\!\begin{tabular}{c}User\\centric\end{tabular}\!\!\!\!\!\!\!\!\\
		\hline
	FACOS & \CIRCLE & \CIRCLE & \CIRCLE  &\CIRCLE &\CIRCLE&\CIRCLE&\CIRCLE\\
		\hdashline[1pt/1pt]
        Ref.~\cite{liang2022pdpchain} & \LEFTcircle & \LEFTcircle  & \Circle  &\Circle &\CIRCLE&\CIRCLE&\CIRCLE\\
		\hdashline[1pt/1pt]
         Ref.~\cite{gao2021bsspd} & \LEFTcircle & \Circle & \Circle &\Circle &\Circle&\CIRCLE&\CIRCLE \\
		\hdashline[1pt/1pt]
       Ref.~\cite{grabis2020blockchain} & \LEFTcircle & \Circle & \Circle &\Circle &\Circle&\CIRCLE&\Circle \\
            \hdashline[1pt/1pt]
       Ref.~\cite{cong2021individual} & \LEFTcircle & \Circle & \Circle &\Circle &\Circle&\CIRCLE&\CIRCLE \\
            \hdashline[1pt/1pt]
       Ref.~\cite{wang2018blockchain} & \LEFTcircle & \Circle & \Circle &\Circle &\Circle&\Circle&\CIRCLE \\
		\hline
	\end{tabular} \
	\vspace{3pt}
	\caption{COMPARISON WITH OTHER SCHEMES. \CIRCLE~DENOTES FULL SUPPORT.} 
	\label{tab:comparewithotherpapers}
\vspace{-6pt}
\end{table*}

\begin{figure*}[!ht]
    \centering
    \begin{minipage}[t]{0.49\textwidth}
        \centering
         \begin{subfigure}[t]{0.48\textwidth}
            \centering
            \input{evl/Fig13}
            \caption{Latency for ABE encryption and decryption. (The policy number is 10.)}
            \label{fig:ABEcompare}
        \end{subfigure}
        \hspace{0.5pt}
        \begin{subfigure}[t]{0.48\textwidth}
            \centering
            \input{evl/Fig16}
            \caption{Latency for encryption and decryption for FACOS access control.}
            \label{fig:accesscompare}
        \end{subfigure}
        \caption{Comparison of ABE algorithms and FACOS fine-grained access control cryptographic algorithms.}
        \label{fig:compare1}
    \end{minipage}
    \begin{minipage}[t]{0.49\textwidth}
        \centering
        \begin{subfigure}[t]{0.48\textwidth}
            \centering
            \input{evl/CPWrite}
            \caption{Comparison of total writing time between FACOS and Ref. \cite{liang2022pdpchain} in off-chain storage phase.}
            \label{fig:compareWritetime}
        \end{subfigure}
        \hspace{0.5pt}
        \begin{subfigure}[t]{0.48\textwidth}
            \centering
            \input{evl/CPRead}
            \caption{Comparison of total read time between FACOS and Ref. \cite{liang2022pdpchain} in off-chain read phase.}
            \label{fig:comparereadtime2}
        \end{subfigure}
        \caption{Comparison of total time between FACOS and Ref. \cite{liang2022pdpchain}}
        \label{fig:compare2}
    \end{minipage}
\end{figure*}

Agreement is implied by agreement of the BFT protocol and the decryption consistency of the labeled attribute encryption. If a correct data requester delivers a message $m$ matching its on-chian $txid$, the corresponding write operation in the form of attribute encryption ciphertext was delivered by at least $f + 1$ replicas. Among the $f + 1$ replicas, at least one of them is correct and delivers the write operation. By the Agreement property of the BFT protocol, every correct replica delivers the write operation. Therefore, every correct replica will broadcast its ciphertexts for the write operation to each data requester with the same on-chain $id$. Accordingly, each
data requester will receive $f + 1$ valid ciphertexts and obtain the message $m$. Due to the decryption consistency property of the underlying attribute encryption, we know that every correct data requester with the same on-chain $txid$ will deliver the same message $m$.

We now prove Total Order. In FACOS, data requesters deliver encrypted messages according to the access control list total order assigned by replicas which are totally ordered. (Decryption consistency guarantees if messages in the encrypted form are totally ordered according to the access control list, then messages in plaintext form are delivered in total
order.) Therefore, we just need to ensure that access control will not cause problems. In our system, data requesters will skip empty messages for the write phase which data requesters are not authorized to see. Namely, once each data requester receives $f + 1$ empty messages with some sequence number $s_n$ from off-chain replicas, it skips $s_n$ and proceeds to deliver other non-empty messages in sequence number order. This way, message total order is achieved for all data requesters.

We first prove the first part of Liveness. If a data owner is correct and submits an encrypted message $m$ with its access control list matching an on-chain $txid$, then replicas will send ciphertexts to all data requesters with the on-chain $txid$. All data requesters will deliver valid ciphertexts in sequence number order and deliver the corresponding message $m$. All data requesters will then deliver valid ciphertexts for the access control list after all messages with smaller per-message sequence numbers are delivered or empty messages with smaller sequence numbers are skipped. Empty messages will be non-blocking, and Liveness is achieved. If a data requester issues an on-chain $txid$, replicas will send all matching messages and all data requesters will receive all messages matching the access control list.

Authentication follows from channel authentication between data owners and replicas and between replicas and data requesters. 

Uniqueness easily follows from the fact that each data requester maintains a log of received messages and delivers each of them in sequence number order and by once. Confidentiality and access control follow from the chosen-ciphertext security of the attribute encryption.

\subsection{Property analysis}

\textbf{1) Distributed denial of service (DDoS) attacks}

In our protocol, the off-chain BFT storage utilizes the asynchronous BFT protocol which does not rely on timing assumption and is robust against DDoS attacks. Simoutiously, for the on-chain phase, the client is only permitted to initiate access requests after obtaining the authorization authorized by the blockchain network. Subsequently, when the client initiates an access request towards the blockchain, the blockchain can verify the transaction's validity. Unauthorized clients, even if they attempt to initiate access requests, will not have obtained the necessary access rights. As a result, the blockchain network refrains from processing their requests. This effectively prevents unauthorized clients from launching an excessive volume of requests toward blockchain, thereby mitigating the risk of system paralysis.

\textbf{2) Data integrity}

When the data requester reads the ciphertexts with the $h$, it can use the corresponding access control method to decrypt the ciphertexts and get $m$. The data requester could check $h$ equals to $hash$($m$), if it holds, the data integrity is preserved; otherwise, it aborts.

\textbf{3) Non-repudiation attack}

In our protocol, when clients initiate authorized access requests, they invoke a smart contract for permission verification, which results in the creation of a trustworthy and immutable access log on the blockchain. Any unlawful access attempts by malicious clients are permanently recorded on the blockchain, making it impossible for them to deny their actions.

%% file: evl/Fig13.tex
\definecolor{Color01}{HTML}{A30141}
\definecolor{Color02}{HTML}{EC6043}
\definecolor{Color03}{HTML}{FCCA78}
\definecolor{Color04}{HTML}{F9FCB7}
\definecolor{Color05}{HTML}{aadda5}
\definecolor{Color06}{HTML}{3893b8}
\definecolor{Color07}{HTML}{4966ae}

\begin{tikzpicture}[scale=0.4]
\begin{axis}[
    every axis plot post/.style={/pgf/number format/fixed},
    ybar,
    bar width=11pt,
    x=2.5cm,
    y=0.0192cm,
    ymin=0,
    axis on top,
    ylabel={Latency (ms)},
    ymax=249,
    xtick=data,
    axis line style={-latex},
    enlarge x limits=0.4,
    symbolic x coords={Generation,Encryption,Decryption},
    visualization depends on=rawy\as\rawy, 
    after end axis/.code={ 
            \draw [ultra thick, white, decoration={snake, amplitude=1pt}, decorate] (rel axis cs:0,1.0) -- (rel axis cs:1,1.0);
        },
    nodes near coords={%
            \pgfmathprintnumber{\rawy}
        },
    every node near coord/.append style={font=\tiny, color=black},
    axis lines*=left,
    clip=false,
    legend columns=4,
    legend image code/.code={%
      \draw[#1] (0cm,-0.1cm) rectangle (0.4cm,0.1cm);
    },
    legend style={at={(0.05,1)},anchor=north west}
    ]

\addplot+[draw opacity=0, color=Color01!70] coordinates {(Generation,83.30) (Encryption,40.71) (Decryption,14.08)};
\addplot+[draw opacity=0, color=Color02!70] coordinates {(Generation,181.83) (Encryption,71.01) (Decryption,15.83)};
\addplot+[draw opacity=0, color=Color03!70] coordinates {(Generation,38.15) (Encryption,46.79) (Decryption,30.64)};
\addplot+[draw opacity=0, color=Color07!70] coordinates {(Generation,69.15) (Encryption,43.22) (Decryption,52.89)};

\legend{AC'17,CGW'15,Waters'11,BSW'07}
\end{axis}
\end{tikzpicture}

%% file: evl/Fig16.tex
\definecolor{Color01}{HTML}{A30141}
\definecolor{Color02}{HTML}{EC6043}
\definecolor{Color03}{HTML}{FCCA78}
\definecolor{Color04}{HTML}{F9FCB7}
\definecolor{Color05}{HTML}{aadda5}
\definecolor{Color06}{HTML}{3893b8}
\definecolor{Color07}{HTML}{4966ae}

\begin{tikzpicture}[scale=0.4]
\begin{axis}[
    every axis plot post/.style={/pgf/number format/fixed},
    ybar,
    bar width=11pt,
    x=2.9cm,
    y=0.43cm,
    ymode=log,
    axis on top,
    ylabel={Latency (ms)},
    ymax=3000,
    xtick=data,
    axis line style={-latex},
    enlarge x limits=1,
    symbolic x coords={Encryption,Decryption},
    visualization depends on=rawy\as\rawy,
    after end axis/.code={
            \draw [ultra thick, white, decoration={snake, amplitude=1pt}, decorate] (rel axis cs:0,1.0) -- (rel axis cs:1,1.0);
        },
    nodes near coords align=above,
    nodes near coords={
            \pgfmathprintnumber{\rawy}
        },
    every node near coord/.append style={font=\tiny, color=black},
    axis lines*=left,
    clip=false,
    legend columns=3,
    legend image code/.code={
      \draw[#1] (0cm,-0.1cm) rectangle (0.4cm,0.1cm);
    },
    legend style={at={(0.05,1)},anchor=north west}
    ]
\addplot+[draw opacity=0, color=Color01!70] coordinates {(Encryption,148.63) (Decryption,116.91)};
\addplot+[draw opacity=0, color=Color02!70] coordinates {(Encryption,0.16) (Decryption,0.13)};
\addplot+[draw opacity=0, color=Color03!70] coordinates {(Encryption,1.13) (Decryption,2.62)};

\legend{ABE module, BE module, TE module}
\end{axis}
\end{tikzpicture}

%% file: evl/CPWrite.tex
\begin{tikzpicture}[scale=0.45]
\begin{axis}[
    xlabel={Data (KB)},
    ylabel={Latency (s)},
    x=0.55cm,
    y=1.44cm,
    xmin=1, xmax=13,
    ymin=0, ymax=2.6,
    xtick={0, 2.5, 5, 7.5, 10, 12.5},
    xticklabels={0, 1, 2, 4, 8, 16},
    ytick={0.4, 0.8, 1.2, 1.6, 2, 2.4},
    legend pos=north west,
    axis lines=left,
    ymajorgrids=true,
    grid style=dashed,
    legend style={
        at={(0.39, 1)},
        anchor=north,
    },
    legend cell align=left,
    legend entries={
        {Ref. \cite{liang2022pdpchain}},
        {FACOS with ABE module},
        {FACOS with BE module},
        {FACOS with TE module}
    },
]

\addplot[color=blue, mark=square]
    coordinates {
        (2.5, 0.4874198437)
        (5, 0.603269577)
        (7.5, 0.8245818615)
        (10, 1.25754714)
        (12.5, 2.082303524)
    };

\addplot[color=red, mark=square]
    coordinates {
        (2.5, 0.4908121228)
        (5, 0.4748963118)
        (7.5, 0.4747780561)
        (10, 0.4797903299)
        (12.5, 0.4759005308)
    };

\addplot[color=green, mark=square]
    coordinates {
        (2.5, 0.2263704538)
        (5, 0.2577457428)
        (7.5, 0.2282611728)
        (10, 0.2447602153)
        (12.5, 0.241712749)
    };

\addplot[color=yellow, mark=square]
    coordinates {
        (2.5, 0.2839398384)
        (5, 0.2623093128)
        (7.5, 0.2721028924)
        (10, 0.2821104527)
        (12.5, 0.2780100107)
    };
    
\end{axis}
\end{tikzpicture}

%% file: evl/CPRead.tex
\begin{tikzpicture}[scale=0.45]
\begin{axis}[
    xlabel={Data (KB)},
    ylabel={Latency (s)},
    x=0.55cm,
    y=2.9cm,
    xmin=1, xmax=13,
    ymin=0, ymax=1.3,
    xtick={0, 2.5, 5, 7.5, 10, 12.5},
    xticklabels={0, 1, 2, 4, 8, 16},
    ytick={0.2, 0.4, 0.6, 0.8, 1.0, 1.2},
    legend pos=north west,
    axis lines=left,
    ymajorgrids=true,
    grid style=dashed,
    legend style={
        at={(0.39, 1)},
        anchor=north
    },
    legend cell align=left,
    legend entries={
        {Ref. \cite{liang2022pdpchain}},
        {FACOS with ABE module},
        {FACOS with BE module},
        {FACOS with TE module}
    },
]

\addplot[color=blue, mark=square]
    coordinates {
        (2.5, 0.2510178089)
        (5, 0.274595499)
        (7.5, 0.4156870842)
        (10, 0.544311285)
        (12.5, 0.7922270298)
    };

\addplot[color=red, mark=square]
    coordinates {
        (2.5, 0.4634122849)
        (5, 0.4395526648)
        (7.5, 0.4670348763)
        (10, 0.4305383563)
        (12.5, 0.4442958236)
    };

\addplot[color=green, mark=square]
    coordinates {
        (2.5, 0.2516723871)
        (5, 0.2460859418)
        (7.5, 0.2228155136)
        (10, 0.2422247529)
        (12.5, 0.2301574945)
    };

\addplot[color=yellow, mark=square]
    coordinates {
        (2.5, 0.2465467453)
        (5, 0.218190968)
        (7.5, 0.2318544984)
        (10, 0.2428066134)
        (12.5, 0.2618028522)
    };
    
\end{axis}
\end{tikzpicture}

%% file: tex/evaluation.tex
\section{Implementation and evaluation}

The entire library includes 10,000 lines of Python codes, among which 500 lines of code are used for evaluation. Our off-chain BFT storage is based on the Dumbo \cite{guo2020dumbo} library. We used the Charm Python library~\cite{akinyele2013charm}. To better understand the overhead incurred by the attribute encryption scheme cryptography, we also directly use Charm's PBC library to implement an attribute encryption scheme, AC'17 \cite{agrawal2017fame}, with SS512 group. This group has 80-bit security. The trusted verifier utilizes the Gramine framework which is based on SGX. The entire application includes 86 lines of Python code and 109 lines of manifest content. The loader's entry point is set ``gramine.libos", thus the unmodified application can run in the SGX. The function of edmm is set to unavailable. The enclave size is set to 1G, and the max threads are 32 in the enclave. Due to the experiment in \cite{liang2022pdpchain} not disclosing the algorithm and library used, we employed AC'17 as the access control scheme. We also implemented \cite{liang2022pdpchain} with the homomorphic encryption library (https://github.com/data61/python-paillier) while also incorporating IPFS (https://ipfs.tech/) for off-chain storage.
\begin{figure}[!ht]
    \centering
    \includegraphics[width=0.85\linewidth]{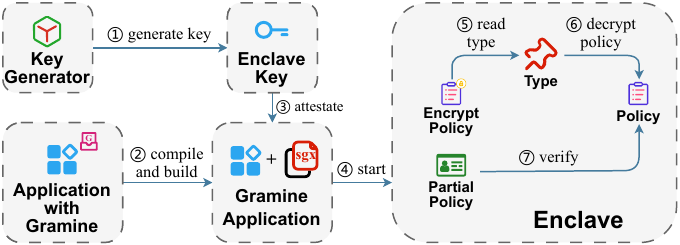}
    \caption{SGX workflow.}
    \label{fig:enter-label}
\end{figure}

Fig. \ref{fig:enter-label} shows the whole process of SGX. \textcircled{\scriptsize 1} Prepare a signing key by key generator, it will generate an RSA 3072 key suitable for signing SGX enclaves and it usually stores at ``\$HOME/.config/gramine/enclave-key.pem". \textcircled{\scriptsize 2} Compile and build the application with Gramine, files running in the trust section are defined in ``python.manifest.template". \textcircled{\scriptsize 3} Attestation the enclaves by key in \textcircled{\scriptsize 1}, the enclave must be certified to be available to protect the code in the enclave. \textcircled{\scriptsize 4} Start the application run in the enclave. In the enclave phase, \textcircled{\scriptsize 5} read the type from the encrypt policy, it is the encrypt method of the Policy by the data owner. \textcircled{\scriptsize 6} Decrypt policy by the type from \textcircled{\scriptsize 5}. \textcircled{\scriptsize 7} Verify the partial attribute of the data requester.

\subsection{Experimental Settings}
We conducted our evaluation on both Alibaba Cloud and Tencent Cloud, utilizing up to 11 virtual machines (VMs). Each VM is equipped with two vCPUs and 4GB of memory, and they all operate on Ubuntu 20.04. For the WAN configuration, VMs are uniformly distributed across different regions. Our assessment of the protocols was based on varying network sizes and batch sizes. For the SGX deployment on the Alibaba server, the memory allocation consisted of a total of 8GB, with 4GB dedicated as physical memory and the remaining 4GB used as encrypted memory. For the blockchain network, we use Hyperledger Fabric V2.3. We deployed Fabric on a single server, and the consensus algorithm used is the RAFT protocol. For the Chaincode, we only simulated the read and write functions within 300 lines of codes in Golang. We use $f$ to represent the network size, and the total number of off-chain replicas is $n = 3f+1$. $K$ denotes the total number of data to be agreed upon. We set $f=1$, $n=4$, $B=40$, $b=10$, and $K=1000$.

We compare FACOS with schemes \cite{liang2022pdpchain,gao2021bsspd,grabis2020blockchain,cong2021individual,wang2018blockchain} in terms of fine-grained access control, practical and scalable, formal proof, Byzantine off-chain storage, data security, data privacy protection, and user-centric, respectively, and the results of the security analysis are shown in Table \ref{tab:comparewithotherpapers}. Compared to other schemes in different aspects, such as data security, data privacy protection, user-centric, fine-grained access control of data, and practical and scalable, the proposed scheme has better security performance and is more suitable for
trusted storage and secure sharing of personal privacy data in the consortium blockchain.
\begin{figure}[!ht]
  \centering
  \begin{subfigure}{0.44\textwidth}
    \centering
    \includegraphics[width=\textwidth]{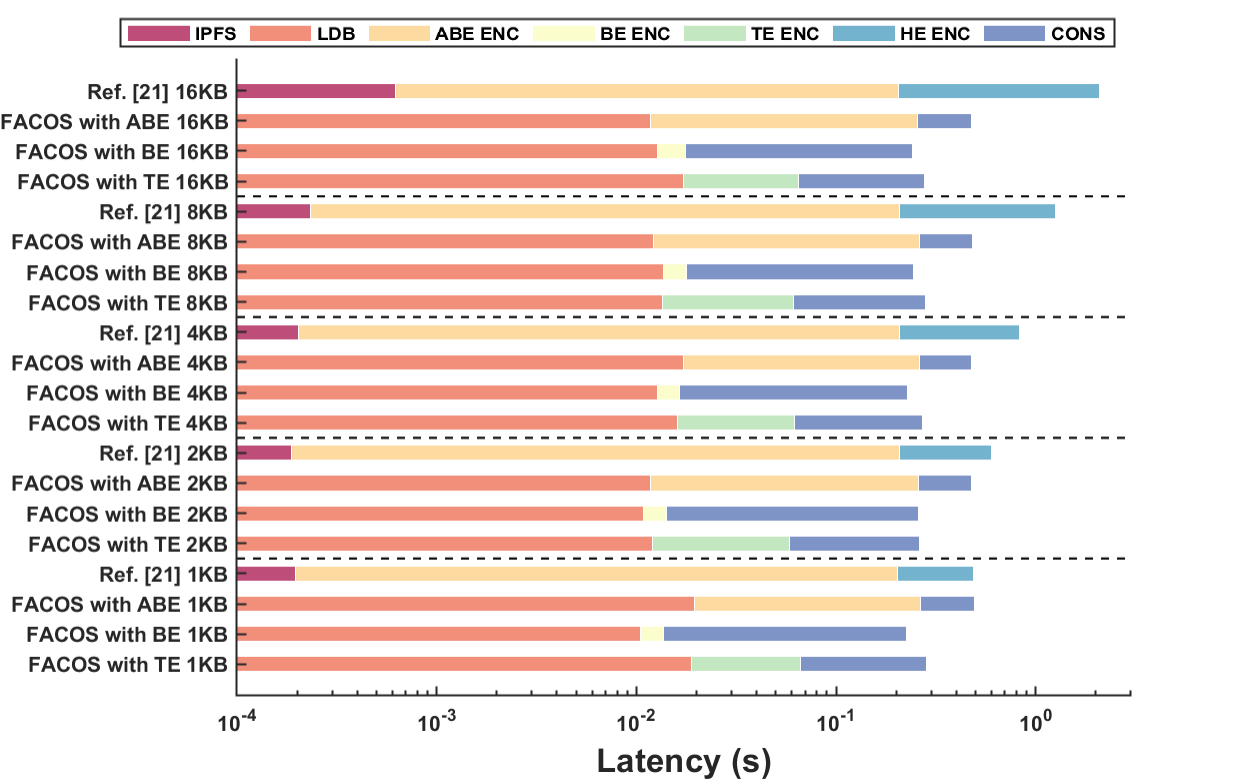}
    \caption{Presenting a phased breakdown of the time spent in each stage for off-chain writing in FACOS and \cite{liang2022pdpchain}.}
    \label{fig:comparephaseswritetime}
  \end{subfigure}
  \begin{subfigure}{0.44\textwidth}
    \centering
    \includegraphics[width=\textwidth]{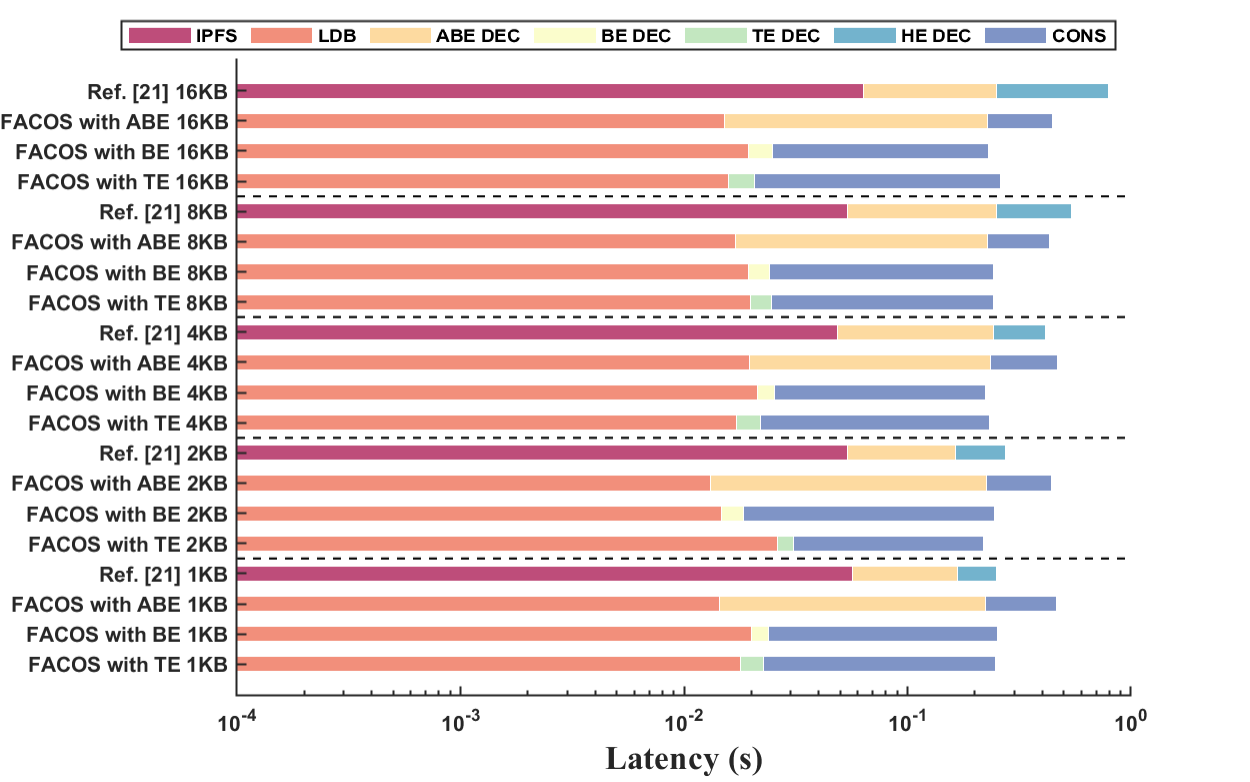}
    \caption{Presenting a phased breakdown of the time spent in each stage for off-chain reading in FACOS and \cite{liang2022pdpchain}.}
    \label{fig:comparephasesreadtime}
  \end{subfigure}
  \caption{FACOS is compared with Ref. \cite{liang2022pdpchain}.}
  \label{fig:compare}
\end{figure}
In Fig. \ref{fig:compare2} and Fig. \ref{fig:compare}, ee compare  \cite{liang2022pdpchain} with FACOS from different perspectives. Due to the adoption of Hyperledger Fabric on-chain, we do not describe the on-chain time but mainly focus on the time for writing to and reading from the off-chain. Figures \ref{fig:compareWritetime} and \ref{fig:comparereadtime2} present the overall time comparison between \cite{liang2022pdpchain} and our three approaches. From the results, it can be observed that the latency of \cite{liang2022pdpchain} increases significantly as the message size increases. Figures \ref{fig:comparephaseswritetime} and \ref{fig:comparephasesreadtime} demonstrate the differences between \cite{liang2022pdpchain} and the FACOS approach in a segmented form, providing an intuitive comparison of the different algorithms and frameworks adopted by \cite{liang2022pdpchain} and FACOS. Through analysis of experimental results, when the data volume reaches 16kb, the total time for write and read off-chain in FACOS is only 77.14\% and 43.94\% of Ref. \cite{liang2022pdpchain}. Analysis of Fig. \ref{fig:compare2} and \ref{fig:compare} show that large data primarily utilizes IPFS for off-chain storage, which offers faster rates. However, with an increase in message volume, the time for homomorphic encryption keeps growing, leading to an extended time in \cite{liang2022pdpchain}. Meanwhile, although FACOS employs asynchronous Byzantine fault tolerance for off-chain storage, the performance is not inferior to the literature due to the parallelization and batch processing capabilities utilized in asynchronous Byzantine storage.
\begin{figure*}[!ht]
  \centering

  \begin{subfigure}{0.3\textwidth}
    \centering
    \input{evl/DP-a}
    \caption{Latency for ABE encryption and decryption in different amounts of attributes. (The ABE scheme uses AC'17, and the number of policy attributes is $2^i(0\leq i\leq9)$.)}
    \label{fig:policysubfigA}
  \end{subfigure}
  \hspace{0.5pt}
  \begin{subfigure}{0.3\textwidth}
    \centering
    \input{evl/DP-b}
    \caption{Latency for BE encryption and decryption in different amounts of clients. (The number of leaf nodes in broadcast encryption is $2^i$, and the number of users deleted is $2^{i-1}-1(11\leq i\leq20)$.)}
    \label{fig:policysubfigB}
  \end{subfigure}
  \hspace{0.5pt}
  \begin{subfigure}{0.3\textwidth}
    \centering
    \input{evl/DP-c}
    \caption{Latency for TE encryption and decryption in different sizes of threshold. (For threshold encryption, the threshold is $f+1$ and the total number of clients is $3f+1$.)\\}
    \label{fig:policysubfigC}
  \end{subfigure}

  \caption{Three access control methods. (The size of the data is 250 bytes.)}
  \label{fig:differentPolicies-b}
\end{figure*}
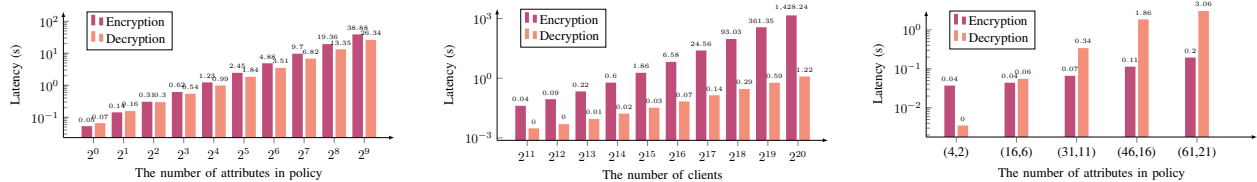 

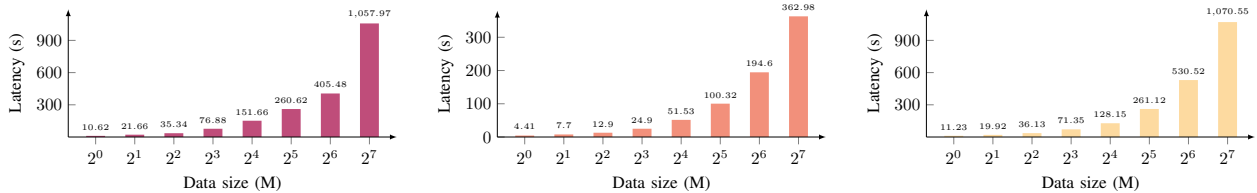
\begin{figure*}[!ht]
  \centering

  \begin{subfigure}{0.3\textwidth}
    \centering
    \input{evl/Fig14-a}
    \caption{Latency for ABE in different amounts of attributes. (The ABE has a policy with 3 attributes.)\ \ \ }
    \label{fig:filesizesubfigA}
  \end{subfigure}
  \hspace{0.5pt}
  \begin{subfigure}{0.3\textwidth}
    \centering
    \input{evl/Fig14-b}
    \caption{Latency for BE in different amounts of clients. (For BE, the total number of users is 4, with 1 being revoked.)}
    \label{fig:filesizesubfigB}
  \end{subfigure}
  \hspace{0.5pt}
  \begin{subfigure}{0.3\textwidth}
    \centering
    \input{evl/Fig14-c}
    \caption{Latency for TE in different sizes of threshold. (For TE, the total is 4 with a threshold of 2.)}
    \label{fig:filesizesubfigC}
  \end{subfigure}
  
  \caption{Write phase with different file sizes. (The data size is $2^i$ M$~(0\leq i\leq7)$.)}
  \label{fig:filesize}
\end{figure*}
\subsection{ABE schemes}
We implement four ABE schemes, namely BSW'07 \cite{bethencourt2007ciphertext}, Waters'11 \cite{waters2011ciphertext}, CGW'15 \cite{chen2015improved}, and AC'17 \cite{agrawal2017fame}. We compare these schemes in three phases, key generation, encryption, and decryption, and we only show their average latency in Fig. \ref{fig:accesscompare} (the number of attribute lists is 10.). During the key generation stage, Waters'11 performs the best, with a latency of 38.15 ms. During the encryption phase, AC'17 is the top performer with a latency of 40.71 ms. In the decryption phase, AC'17 stands out with the best performance. In conclusion, considering efficiency concerns, we adopt AC'17 as the ABE scheme for FACOS.

\subsection{Access control methods}
\noindent We have three access control methods (attribute encryption   \cite{agrawal2017fame}, broadcast encryption \cite{fiat1994broadcast}, and threshold encryption \cite{shoup1998securing}). When evaluating these three access control schemes, we conducted 1000 tests for each. We calculated the encryption and decryption times by taking the average, and the result is shown in Fig. \ref{fig:accesscompare} (the number of attribute lists for ABE is 3. For BE, the total number of clients is 4, with 1 being revoked. For TE, the total number of clients is 4 with a threshold of 2).

In summary, the performance differences among these three encryption methods align precisely with their underlying cryptographic principles shown in Table \ref{tab:detailedthreemethods}. Attribute encryption, which relies on pairing operations, exhibits relatively high encryption and decryption latencies. Broadcast encryption, grounded in symmetric cryptography, showcases outstanding performance with remarkably low encryption and decryption latencies. Threshold encryption, based on elliptic curve cryptography, displays a performance gap between the encryption and decryption phases. This indicates a trade-off between security and efficiency.

\subsection{Access control policy}

For attribute encryption, shown in Fig. \ref{fig:policysubfigA}, as the number of attributes in the policy increases, encryption latency also shows an upward trend. We tested encryption and decryption times for the threshold encryption access control scheme by varying the threshold size $f$ and the total number of clients $n$, where the threshold is $f+1$. The test result is shown in Fig. \ref{fig:policysubfigC}. The results align with the theoretical basis. As the complexity of attribute policies increases, both the height of the broadcast encryption tree (associated with the growth in the number of clients) and the threshold encryption threshold (associated with the rise in total clients) also increase. This leads to a corresponding increase in latency in the experimental outcomes.

\subsection{File sizes}
By varying the file sizes and using the attribute encryption scheme for writeing a single file, we conducted multiple tests to calculate the average protocol runtime. The experimental results in Fig. \ref{fig:filesizesubfigA} demonstrate that as the file size increases, the runtime correspondingly grows. As the file size increases, latency is primarily observed during the off-chain consensus phase. The consensus phase necessitates the initial fragmentation of the entire message and the broadcast of each fragment. The broadcast in the off-chain consensus requires at least three stages, significantly amplifying the delay. It's worth noting that broadcast encryption takes less time because it eventually uses stream cipher encryption. The data encoded through this method occupies less bandwidth, leading to a shorter time spent during the consensus phase. We conducted the same simulation experiment to verify this perspective. When encrypting the data using the AES encryption algorithm, we found that the time taken for broadcast encryption is roughly the same as the time spent using the other two encryption methods.

\begin{figure*}[!ht]
  \centering
  \begin{subfigure}{0.3\textwidth}
    \centering
    \input{evl/UAD-d}
    \caption{Write with ABE module.}
    \label{fig:uploadsubfigA}
  \end{subfigure}
  \hspace{0.5pt}
  \begin{subfigure}{0.3\textwidth}
    \centering
    \input{evl/UAD-e}
    \caption{Write with BE module.}
    \label{fig:uploadsubfigB}
  \end{subfigure}
  \hspace{0.5pt}
  \begin{subfigure}{0.3\textwidth}
    \centering
    \input{evl/UAD-f}
    \caption{Write with TE module.}
    \label{fig:uploadsubfigC}
  \end{subfigure}
    \hspace{0.5pt}
  \begin{subfigure}{0.3\textwidth}
    \centering
    \input{evl/UAD-a}
    \caption{Read with ABE module.}
    \label{fig:downloadsubfigD}
  \end{subfigure}
     \hspace{0.5pt}
  \begin{subfigure}{0.3\textwidth}
    \centering
    \input{evl/UAD-b}
    \caption{Read with BE module.}
    \label{fig:downloadsubfigE}
  \end{subfigure}
     \hspace{0.5pt}
  \begin{subfigure}{0.3\textwidth}
    \centering
    \input{evl/UAD-c}
    \caption{Read with TE module.}
    \label{fig:downloadsubfigF}
  \end{subfigure}
  \caption{Write and read phase. (The attribute encryption employs a policy comprising 3 attributes. In the broadcast encryption, there are 4 clients in total, with 1 of them being revoked. In the threshold encryption, there are 4 clients with a set threshold of 2. The data size amounts to 250 bytes, and the client count is represented by $n$ (where $n$ can be values such as 1, 10, 100, 500, 1000, or 2000).}
  \label{fig:upAndDownload}
\end{figure*}
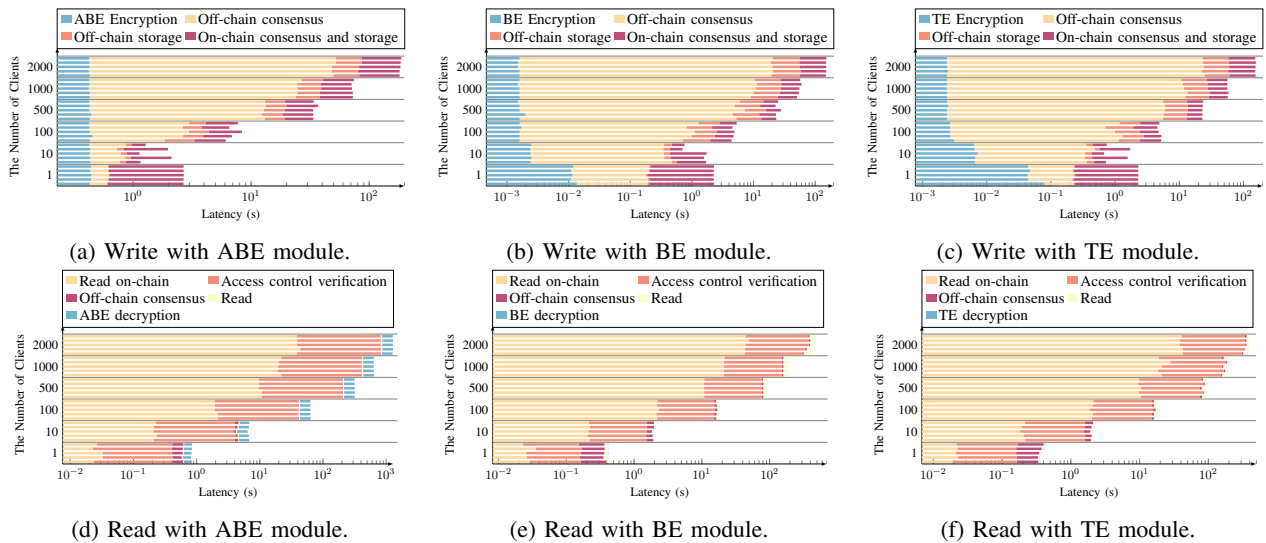

\subsection{Efficiency analyze}
For simplicity, we take the ABE access control method as an example. 

The write phase consists of four steps: client encryption, off-chain data consensus, off-chain data storage, and on-chain storage. Clients are simulated by using multiple servers and threads. In Fig. \ref{fig:upAndDownload}, the y-axis is in a logarithmic scale, indicating the number of clients (ranging from 1 to 2000), and the x-axis represents latency.

For the writing phase, multiple client simulations are performed using a multi-server, multi-threaded approach. The encryption process is executed concurrently, meaning that even with an increase in the number of clients, the runtime of the client encryption phase remains relatively consistent. After multiple off-chain consensus evaluations, we optimized the data volume for each consensus round. In each round, servers take out up to 10 pieces of data for consensus, and the maximum number of data items that can be processed in a single consensus round is $10*(n-f)$. As the number of clients increases, it leads to an increase in the number of consensus rounds required and, consequently, in the time needed. Off-chain consensus and off-chain storage run concurrently. However, off-chain storage employs LevelDB, which cannot achieve concurrent storage. The runtime of off-chain storage is related to the amount of data being stored, and as the data volume increases, the runtime also extends. In other words, as the number of clients increases, off-chain storage time exhibits linear growth. For on-chain storage, we implemented the bundling of multiple pieces of information for simultaneous blockchain transactions. When the number of clients is low, the runtime of on-chain storage remains relatively stable. As the number of clients increases, the on-chain storage time also increases. However, with a significant increase in the number of clients, the average time for on-chain transactions tends to stabilize, aligning with real-world scenarios.

During the Read phase, we simulate the process only using a single thread. When reading data from the blockchain, the runtime of on-chain reading increases linearly with the growth in the number of users. The runtime of verification and decryption depends on the method of access control. In the case of implementing verification in SGX, the time spent on verification is related to the complexity of access control. The access control scheme based on attribute encryption has the longest verification runtime, while the threshold encryption access control scheme has the shortest. Regarding decryption, the results align with the earlier decryption tests. The access control scheme based on attribute encryption exhibits longer runtimes for both verification and decryption, while the broadcast encryption access control scheme has shorter runtimes. For off-chain data reading, consensus is initially reached off-chain for reading commands and indexes. Then, data is read from the database and returned to the user. The data for consensus regarding reading commands and indexes is relatively small, resulting in fast off-chain consensus. LevelDB is used for the database, ensuring quick data retrieval. User decryption follows the same patterns observed in the decryption tests mentioned earlier.

Upon comparison, it was found that the system spends more time during the read phase. The primary reason is that when the protocol runs the read operation, it only implements a single thread. The runtime during the three stages of read from the Fabric, SGX access comparison, and user decryption of $c_h$ (off-chain index) show a linear increase. However, off-chain consensus and data read use multi-threading, so the basic time remains unchanged compared with the write phase.

%% file: evl/DP-a.tex
\definecolor{Color01}{HTML}{A30141}
\definecolor{Color02}{HTML}{EC6043}
\definecolor{Color03}{HTML}{FCCA78}
\definecolor{Color04}{HTML}{F9FCB7}
\definecolor{Color05}{HTML}{aadda5}
\definecolor{Color06}{HTML}{3893b8}
\definecolor{Color07}{HTML}{4966ae}

\begin{tikzpicture}[scale=0.5]
\begin{axis}[
    every axis plot post/.style={/pgf/number format/fixed},
    ybar,
    bar width=8pt,
    x=0.8cm,
    y=0.37cm,
    axis on top,
    ylabel={Latency (s)},
    xmax=11,
    ymax=200,
    ymode=log,
    axis line style={-latex},
    xlabel={The number of attributes in policy},
    xtick={1,2,3,4,5,6,7,8,9,10},
    xticklabels={$2^0$,$2^1$,$2^2$,$2^3$,$2^4$,$2^5$,$2^6$,$2^7$,$2^8$,$2^9$},
    visualization depends on=rawy\as\rawy,
    nodes near coords={
        \pgfmathprintnumber{\rawy}
    },
    nodes near coords align=above,
    every node near coord/.append style={
        font=\tiny, 
        text=black,
    },
    axis lines*=left,
    clip=false,
    legend style={
        at={(0.2,1)},
        anchor=north
    },
    legend cell align=left,
    legend entries={
            {Encryption},
            {Decryption}
        },
]
    \addplot+[draw opacity=0,color=Color01!70] coordinates {(1,0.0530) (2,0.1429) (3,0.3068) (4,0.6188) (5,1.2328) (6,2.4495) (7,4.8763) (8,9.6984) (9,19.3571) (10,38.8782)};
    \addplot+[draw opacity=0,color=Color02!70] coordinates {(1,0.0657) (2,0.1566) (3,0.2984) (4,0.5415) (5,0.9877) (6,1.8423) (7,3.5139) (8,6.8165) (9,13.3496) (10,26.3436)};

\end{axis}
\end{tikzpicture}

%% file: evl/DP-b.tex
\definecolor{Color01}{HTML}{A30141}
\definecolor{Color02}{HTML}{EC6043}
\definecolor{Color03}{HTML}{FCCA78}
\definecolor{Color04}{HTML}{F9FCB7}
\definecolor{Color05}{HTML}{aadda5}
\definecolor{Color06}{HTML}{3893b8}
\definecolor{Color07}{HTML}{4966ae}

\begin{tikzpicture}[scale=0.5]
\begin{axis}[
    every axis plot post/.style={/pgf/number format/fixed},
    ybar,
    bar width=8pt,
    x=0.8cm,
    y=0.23cm,
    axis on top,
    ylabel={Latency (s)},
    xmax=11,
    ymax=3000,
    ymode=log,
    xlabel={The number of clients},
    axis line style={-latex},
    xtick={1,2,3,4,5,6,7,8,9,10},
    xticklabels={$2^{11}$,$2^{12}$,$2^{13}$,$2^{14}$,$2^{15}$,$2^{16}$,$2^{17}$,$2^{18}$,$2^{19}$,$2^{20}$},
    visualization depends on=rawy\as\rawy, 
    nodes near coords={%
            \pgfmathprintnumber{\rawy}
        },
    every node near coord/.append style={font=\tiny, color=black},
    nodes near coords align=above,
    axis lines*=left,
    clip=false,
    legend style={
        at={(0.2,1)},
        anchor=north
    },
    legend cell align=left,
    legend entries={
        {Encryption},
        {Decryption}
    },
]
\addplot+[draw opacity=0,color=Color01!70] coordinates {(1,0.0408) (2,0.0879) (3,0.2155) (4,0.5983) (5,1.8644) (6,6.5785) (7,24.5580) (8,93.0274) (9,361.3485) (10,1428.2422)};
\addplot+[draw opacity=0,color=Color02!70] coordinates {(1,0.0030) (2,0.0049) (3,0.0088) (4,0.0166) (5,0.0327) (6,0.0677) (7,0.1387) (8,0.2864) (9,0.5882) (10,1.2170)};

\end{axis}
\end{tikzpicture}
 

%% file: evl/DP-c.tex
\definecolor{Color01}{HTML}{A30141}
\definecolor{Color02}{HTML}{EC6043}
\definecolor{Color03}{HTML}{FCCA78}
\definecolor{Color04}{HTML}{F9FCB7}
\definecolor{Color05}{HTML}{aadda5}
\definecolor{Color06}{HTML}{3893b8}
\definecolor{Color07}{HTML}{4966ae}

\begin{tikzpicture}[scale=0.5]
\begin{axis}[
    every axis plot post/.style={/pgf/number format/fixed},
    ybar,
    bar width=8pt,
    x=1.6cm,
    y=0.45cm,
    ymode=log,
    axis on top,
    ylabel={Latency (s)},
    xlabel={The number of attributes in policy},
    xmax=6,
    ymax=3.1,
    axis line style={-latex},
    xtick={1,2,3,4,5},
    xticklabels={(4,2),(16,6),(31,11),(46,16),(61,21)},
    visualization depends on=rawy\as\rawy,
    nodes near coords={%
            \pgfmathprintnumber{\rawy}
        },
        every node near coord/.append style={font=\tiny, color=black},
        nodes near coords align=above,
    axis lines*=left,
    clip=false,
    legend style={
        at={(0.2,1)},
        anchor=north
    },
    legend cell align=left,
    legend entries={
        {Encryption},
        {Decryption}
        },
    ]
    \addplot+[draw opacity=0,color=Color01!70] coordinates {(1,0.0372) (2,0.0440) (3,0.0663) (4,0.1138) (5,0.1955)};
    \addplot+[draw opacity=0,color=Color02!70] coordinates {(1,0.0035) (2,0.0554) (3,0.3413) (4,1.857) (5,3.0602)};

\end{axis}
\end{tikzpicture}
 

%% file: evl/Fig14-a.tex
\definecolor{Color01}{HTML}{A30141}
\definecolor{Color02}{HTML}{EC6043}
\definecolor{Color03}{HTML}{FCCA78}
\definecolor{Color04}{HTML}{F9FCB7}
\definecolor{Color05}{HTML}{aadda5}
\definecolor{Color06}{HTML}{3893b8}
\definecolor{Color07}{HTML}{4966ae}

\begin{tikzpicture}[scale=0.65]
\begin{axis}[
    every axis plot post/.style={/pgf/number format/fixed},
    ybar,
    bar width=11pt,
    y=0.0022cm,
    x=0.8cm,
    axis line style={-latex},
    ymin=0,
    axis on top,
    ylabel={Latency (s)},
    ymax=1190,
    xlabel={Data size (M)},
    xtick={1,2,3,4,5,6,7,8},
    ytick={300, 600, 900},
    xticklabels={$2^0$,$2^1$,$2^2$,$2^3$,$2^4$,$2^5$,$2^6$,$2^7$},
    visualization depends on=rawy\as\rawy, 
    nodes near coords={%
            \pgfmathprintnumber{\rawy}
        },
        every node near coord/.append style={font=\tiny, color=black},
    axis lines*=left,
    clip=false,
    ]
\addplot+[draw opacity=0,color=Color01!70] coordinates {(1,10.62) (2,21.66) (3,35.34) (4,76.88) (5,151.66) (6,260.62) (7,405.48) (8,1057.97)};

\end{axis}
\end{tikzpicture}

%% file: evl/Fig14-b.tex
\definecolor{Color01}{HTML}{A30141}
\definecolor{Color02}{HTML}{EC6043}
\definecolor{Color03}{HTML}{FCCA78}
\definecolor{Color04}{HTML}{F9FCB7}
\definecolor{Color05}{HTML}{aadda5}
\definecolor{Color06}{HTML}{3893b8}
\definecolor{Color07}{HTML}{4966ae}

\begin{tikzpicture}[scale=0.65]
\begin{axis}[
    every axis plot post/.style={/pgf/number format/fixed},
    ybar,
    bar width=11pt,
    x=0.8cm,
    y=0.0068cm,
    axis line style={-latex},
    ymin=0,
    axis on top,
    ylabel={Latency (s)},
    ymax=380,
    xlabel={Data size (M)},
    xtick={1,2,3,4,5,6,7,8},
    xticklabels={$2^0$,$2^1$,$2^2$,$2^3$,$2^4$,$2^5$,$2^6$,$2^7$},
    visualization depends on=rawy\as\rawy, 
    nodes near coords={%
            \pgfmathprintnumber{\rawy}
        },
        every node near coord/.append style={font=\tiny, color=black},
    axis lines*=left,
    clip=false,
    ]
\addplot+[draw opacity=0,color=Color02!70] coordinates {(1,4.41) (2,7.70) (3,12.90) (4,24.90) (5,51.53) (6,100.32) (7,194.60) (8,362.98)};

\end{axis}
\end{tikzpicture}

%% file: evl/Fig14-c.tex
\definecolor{Color01}{HTML}{A30141}
\definecolor{Color02}{HTML}{EC6043}
\definecolor{Color03}{HTML}{FCCA78}
\definecolor{Color04}{HTML}{F9FCB7}
\definecolor{Color05}{HTML}{aadda5}
\definecolor{Color06}{HTML}{3893b8}
\definecolor{Color07}{HTML}{4966ae}

\begin{tikzpicture}[scale=0.65]
\begin{axis}[
    every axis plot post/.style={/pgf/number format/fixed},
    ybar,
    bar width=11pt,
    x=0.8cm,
    y=0.0022cm,
    axis line style={-latex},
    ymin=0,
    axis on top,
    ylabel={Latency (s)},
    ymax=1190,
    ytick={300, 600, 900},
    xtick={1,2,3,4,5,6,7,8},
    xlabel={Data size (M)},
    xticklabels={$2^0$,$2^1$,$2^2$,$2^3$,$2^4$,$2^5$,$2^6$,$2^7$},
    visualization depends on=rawy\as\rawy, 
    nodes near coords={%
            \pgfmathprintnumber{\rawy}
        },
        every node near coord/.append style={font=\tiny, color=black},
    axis lines*=left,
    clip=false,
    ]
\addplot+[draw opacity=0,color=Color03!70] coordinates {(1,11.23) (2,19.92) (3,36.13) (4,71.35) (5,128.15) (6,261.12) (7,530.52) (8,1070.55)};

\end{axis}
\end{tikzpicture}

%% file: evl/UAD-d.tex
\definecolor{Color01}{HTML}{A30141}
\definecolor{Color02}{HTML}{EC6043}
\definecolor{Color03}{HTML}{FCCA78}
\definecolor{Color04}{HTML}{F9FCB7}
\definecolor{Color05}{HTML}{aadda5}
\definecolor{Color06}{HTML}{3893b8}
\definecolor{Color07}{HTML}{4966ae}

\begin{tikzpicture}[scale=0.48]
    \begin{axis}[
        xbar stacked,
        axis lines*=left,
        x=1.42cm,
        y=0.12cm,
        axis line style={-latex},
        xmode=log,
        ytick={2.5,7.5,12.5,17.5,22.5,27.5},
        extra y ticks={5,10,15,20,25,30},
        extra y tick labels={,,,,,},
        extra y tick style={major tick length=274pt, thick, black},
        yticklabels={1,10,100,500,1000,2000},
        xtick={1,10,100,1000},
        xticklabels={$10^0$,$10^1$,$10^2$,$10^3$},
        ymin=0,
        ymax=32,
        xmax=200,
        bar width=2pt,
        xlabel={Latency (s)},
        ylabel={The Number of Clients},
        legend columns=2,
        legend style={
            at={(0.5,1.28)},
            anchor=north,
            font=\large,
        },
        legend cell align=left,
        legend entries={
            {ABE Encryption},
            {Off-chain consensus},
            {Off-chain storage},
            {On-chain consensus and storage}
        },
    ]

\addplot+[draw opacity=0,color=Color06!70] coordinates {(0.4288,0.5) (0.4393,1.5) (0.4436,2.5) (0.4338,3.5) (0.4404,4.5) (0.4332,5.5) (0.4355,6.5) (0.4312,7.5) (0.4338,8.5) (0.4277,9.5) (0.4306,10.5) (0.4523,11.5) (0.4297,12.5) (0.4275,13.5) (0.4246,14.5) (0.4301,15.5) (0.4420,16.5) (0.4260,17.5) (0.4299,18.5) (0.4248,19.5) (0.4305,20.5) (0.4297,21.5) (0.4222,22.5) (0.4255,23.5) (0.4266,24.5) (0.4304,25.5) (0.4227,26.5) (0.4285,27.5) (0.4273,28.5) (0.4258,29.5)}; 
\addplot+[draw opacity=0,color=Color03!70] coordinates {(0.1693,0.5) (0.1710,1.5) (0.1778,2.5) (0.1779,3.5) (0.1739,4.5) (0.3374,5.5) (0.4242,6.5) (0.3560,7.5) (0.2938,8.5) (0.4464,9.5) (1.4211,10.5) (2.2033,11.5) (2.5388,12.5) (2.2141,13.5) (2.5588,14.5) (12.6050,15.5) (11.8282,16.5) (12.3808,17.5) (12.8818,18.5) (12.6886,19.5) (23.3888,20.5) (24.2076,21.5) (24.2315,22.5) (24.1844,23.5) (26.1997,24.5) (49.2796,25.5) (47.7538,26.5) (47.5773,27.5) (51.7539,28.5) (51.5605,29.5)}; 
\addplot+[draw opacity=0,color=Color02!70] coordinates {(0.0127,0.5) (0.0093,1.5) (0.0092,2.5) (0.0092,3.5) (0.0085,4.5) (0.0973,5.5) (0.0998,6.5) (0.0960,7.5) (0.1014,8.5) (0.0945,9.5) (1.4561,10.5) (1.2709,11.5) (1.4350,12.5) (1.1629,13.5) (1.1441,14.5) (6.3308,15.5) (6.3581,16.5) (6.3996,17.5) (6.8270,18.5) (6.1589,19.5) (14.3072,20.5) (13.7027,21.5) (14.5986,22.5) (14.0990,23.5) (14.1351,24.5) (32.7173,25.5) (31.7996,26.5) (34.1467,27.5) (34.0212,28.5) (34.1678,29.5)}; 
\addplot+[draw opacity=0,color=Color01!70] coordinates {(2.0670,0.5) (2.0606,1.5) (2.0606,2.5) (2.0636,3.5) (2.0617,4.5) (0.2893,5.5) (1.1603,6.5) (0.2526,7.5) (1.1519,8.5) (0.3107,9.5) (2.7827,10.5) (2.9572,11.5) (3.9808,12.5) (2.7383,13.5) (3.5925,14.5) (14.3589,15.5) (15.0560,16.5) (14.3445,17.5) (16.8812,18.5) (14.6639,19.5) (34.5812,20.5) (34.3340,21.5) (32.3021,22.5) (32.3075,23.5) (33.4446,24.5) (98.6903,25.5) (101.1114,26.5) (101.8874,27.5) (98.7906,28.5) (101.3682,29.5)};

\legend{}

\end{axis}
\end{tikzpicture}

%% file: evl/UAD-e.tex
\definecolor{Color01}{HTML}{A30141}
\definecolor{Color02}{HTML}{EC6043}
\definecolor{Color03}{HTML}{FCCA78}
\definecolor{Color04}{HTML}{F9FCB7}
\definecolor{Color05}{HTML}{aadda5}
\definecolor{Color06}{HTML}{3893b8}
\definecolor{Color07}{HTML}{4966ae}
  
\begin{tikzpicture}[scale=0.48]
    \begin{axis}[
        xbar stacked,
        axis lines*=left,
        x=0.741cm,
        y=0.12cm,
        axis line style={-latex},
        xmode=log,
        ytick={2.5,7.5,12.5,17.5,22.5,27.5},
        extra y ticks={5,10,15,20,25,30},
        extra y tick labels={,,,,,},
        extra y tick style={major tick length=274pt, thick, black},
        yticklabels={1,10,100,500,1000,2000},
        xtick={0.001,0.01,0.1,1,10,100,1000},
        xticklabels={$10^{-3}$,$10^{-2}$,$10^{-1}$,$10^0$,$10^1$,$10^2$,$10^3$},
        ymin=0,
        ymax=32,
        xmax=200,
        bar width=2pt,
        xlabel={Latency (s)},
        ylabel={The Number of Clients},
        legend columns=2,
        legend style={
            at={(0.5,1.28)},
            anchor=north,
            font=\large,
        },
        legend cell align=left,
        legend entries={
            {BE Encryption},
            {Off-chain consensus},
            {Off-chain storage},
            {On-chain consensus and storage}
        },
    ]
    
\addplot+[draw opacity=0,color=Color06!70] coordinates {(0.0135,0.5) (0.0108,1.5) (0.0115,2.5) (0.0109,3.5) (0.0118,4.5) (0.0024,5.5) (0.0025,6.5) (0.0024,7.5) (0.0024,8.5) (0.0025,9.5) (0.0016,10.5) (0.0016,11.5) (0.0016,12.5) (0.0017,13.5) (0.0016,14.5) (0.0016,15.5) (0.0020,16.5) (0.0016,17.5) (0.0016,18.5) (0.0015,19.5) (0.0016,20.5) (0.0016,21.5) (0.0016,22.5) (0.0015,23.5) (0.0016,24.5) (0.0016,25.5) (0.0016,26.5) (0.0015,27.5) (0.0015,28.5) (0.0016,29.5)}; 
\addplot+[draw opacity=0,color=Color03!70] coordinates {(0.1810,0.5) (0.1830,1.5) (0.1682,2.5) (0.1856,3.5) (0.1864,4.5) (0.4638,5.5) (0.3518,6.5) (0.3441,7.5) (0.3404,8.5) (0.3511,9.5) (0.7165,10.5) (1.0056,11.5) (1.1007,12.5) (0.8184,13.5) (1.2742,14.5) (5.2507,15.5) (4.5408,16.5) (7.1522,17.5) (4.9507,18.5) (5.9861,19.5) (8.9807,20.5) (10.1207,21.5) (10.8009,22.5) (11.3623,23.5) (10.4565,24.5) (19.4679,25.5) (20.8380,26.5) (20.1880,27.5) (18.9460,28.5) (20.6104,29.5)}; 
\addplot+[draw opacity=0,color=Color02!70] coordinates {(0.0095,0.5) (0.0094,1.5) (0.0088,2.5) (0.0079,3.5) (0.0133,4.5) (0.1058,5.5) (0.0983,6.5) (0.0957,7.5) (0.0999,8.5) (0.1151,9.5) (1.2740,10.5) (1.3414,11.5) (1.3965,12.5) (1.3080,13.5) (1.5253,14.5) (8.0777,15.5) (7.8447,16.5) (8.7425,17.5) (7.8426,18.5) (8.4180,19.5) (15.7321,20.5) (16.3504,21.5) (16.6930,22.5) (16.8272,23.5) (16.8830,24.5) (36.7231,25.5) (37.1928,26.5) (36.5009,27.5) (36.4649,28.5) (36.2320,29.5)}; 
\addplot+[draw opacity=0,color=Color01!70] coordinates {(2.0790,0.5) (2.0605,1.5) (2.0666,2.5) (2.0662,3.5) (2.0680,4.5) (1.1300,5.5) (1.1225,6.5) (1.2875,7.5) (0.2635,8.5) (0.2887,9.5) (2.4122,10.5) (2.3962,11.5) (2.3898,12.5) (2.4340,13.5) (2.5367,14.5) (10.0693,15.5) (10.7730,16.5) (11.9844,17.5) (10.0441,18.5) (10.7456,19.5) (26.3623,20.5) (28.0118,21.5) (28.2054,22.5) (32.0824,23.5) (29.7663,24.5) (93.2759,25.5) (93.7139,26.5) (94.3995,27.5) (95.1757,28.5) (95.6515,29.5)};

\legend{}

\end{axis}
\end{tikzpicture}

%% file: evl/UAD-f.tex
\definecolor{Color01}{HTML}{A30141}
\definecolor{Color02}{HTML}{EC6043}
\definecolor{Color03}{HTML}{FCCA78}
\definecolor{Color04}{HTML}{F9FCB7}
\definecolor{Color05}{HTML}{aadda5}
\definecolor{Color06}{HTML}{3893b8}
\definecolor{Color07}{HTML}{4966ae}

\begin{tikzpicture}[scale=0.48]
    \begin{axis}[
        xbar stacked,
        axis lines*=left,
        x=0.772cm,
        y=0.12cm,
        axis line style={-latex},
        xmode=log,
        ytick={2.5,7.5,12.5,17.5,22.5,27.5},
        extra y ticks={5,10,15,20,25,30},
        extra y tick labels={,,,,,},
        extra y tick style={major tick length=274pt, thick, black},
        yticklabels={1,10,100,500,1000,2000},
        xtick={0.001,0.01,0.1,1,10,100,1000},
        xticklabels={$10^{-3}$,$10^{-2}$,$10^{-1}$,$10^0$,$10^1$,$10^2$,$10^3$},
        ymin=0,
        ymax=32,
        xmax=200,
        bar width=2pt,
        xlabel={Latency (s)},
        ylabel={The Number of Clients},
        legend columns=2,
        legend style={
            at={(0.5,1.28)},
            anchor=north,
            font=\large,
        },
        legend cell align=left,
        legend entries={
            {TE Encryption},
            {Off-chain consensus},
            {Off-chain storage},
            {On-chain consensus and storage}
        },
    ]
\addplot+[draw opacity=0,color=Color06!70] coordinates {(0.0784,0.5) (0.0441,1.5) (0.0433,2.5) (0.0472,3.5) (0.0439,4.5) (0.0065,5.5) (0.0066,6.5) (0.0072,7.5) (0.0065,8.5) (0.0064,9.5) (0.0031,10.5) (0.0027,11.5) (0.0027,12.5) (0.0027,13.5) (0.0027,14.5) (0.0024,15.5) (0.0024,16.5) (0.0024,17.5) (0.0025,18.5) (0.0024,19.5) (0.0024,20.5) (0.0024,21.5) (0.0024,22.5) (0.0024,23.5) (0.0024,24.5) (0.0024,25.5) (0.0025,26.5) (0.0024,27.5) (0.0024,28.5) (0.0024,29.5)}; 
\addplot+[draw opacity=0,color=Color03!70] coordinates {(0.1497,0.5) (0.1701,1.5) (0.1861,2.5) (0.1698,3.5) (0.1821,4.5) (0.3561,5.5) (0.3276,6.5) (0.3314,7.5) (0.4800,8.5) (0.3484,9.5) (1.1291,10.5) (1.3063,11.5) (0.9911,12.5) (0.6987,13.5) (1.1646,14.5) (5.5679,15.5) (5.6457,16.5) (5.9708,17.5) (5.9353,18.5) (5.5887,19.5) (12.3828,20.5) (13.3617,21.5) (11.7910,22.5) (11.3464,23.5) (10.6801,24.5) (24.5925,25.5) (22.3247,26.5) (24.0537,27.5) (23.2676,28.5) (23.0838,29.5)}; 
\addplot+[draw opacity=0,color=Color02!70] coordinates {(0.0096,0.5) (0.0110,1.5) (0.0093,2.5) (0.0096,3.5) (0.0104,4.5) (0.0990,5.5) (0.1046,6.5) (0.0905,7.5) (0.0949,8.5) (0.0970,9.5) (1.2934,10.5) (1.4499,11.5) (1.4171,12.5) (1.2072,13.5) (1.2912,14.5) (7.2977,15.5) (7.4019,16.5) (7.4960,17.5) (7.5121,18.5) (7.4382,19.5) (16.2556,20.5) (16.7920,21.5) (16.0204,22.5) (16.0015,23.5) (16.3586,24.5) (35.5606,25.5) (35.3549,26.5) (35.7943,27.5) (35.6335,28.5) (34.7613,29.5)}; 
\addplot+[draw opacity=0,color=Color01!70] coordinates {(2.0627,0.5) (2.0648,1.5) (2.0618,2.5) (2.0593,3.5) (2.0590,4.5) (0.2603,5.5) (1.1344,6.5) (0.2444,7.5) (1.1372,8.5) (0.2877,9.5) (2.7750,10.5) (2.4940,11.5) (2.3567,12.5) (2.6420,13.5) (2.4272,14.5) (10.0480,15.5) (9.9216,16.5) (9.4572,17.5) (9.7708,18.5) (10.2935,19.5) (28.3641,20.5) (28.3285,21.5) (28.1701,22.5) (28.6941,23.5) (29.0793,24.5) (94.5119,25.5) (95.2309,26.5) (94.5139,27.5) (96.9907,28.5) (95.2153,29.5)};

\legend{}

\end{axis}
\end{tikzpicture}

%% file: evl/UAD-a.tex
\definecolor{Color01}{HTML}{A30141}
\definecolor{Color02}{HTML}{EC6043}
\definecolor{Color03}{HTML}{FCCA78}
\definecolor{Color04}{HTML}{F9FCB7}
\definecolor{Color05}{HTML}{aadda5}
\definecolor{Color06}{HTML}{3893b8}
\definecolor{Color07}{HTML}{4966ae}

\begin{tikzpicture}[scale=0.48]
    \begin{axis}[
        xbar stacked,
        axis lines*=left,
        x=0.76cm,
        y=0.12cm,
        axis line style={-latex},
        xmode=log,
        ytick={2.5,7.5,12.5,17.5,22.5,27.5},
        extra y ticks={5,10,15,20,25,30},
        extra y tick labels={,,,,,},
        extra y tick style={major tick length=264pt, thick, black},
        yticklabels={1,10,100,500,1000,2000},
        xtick={0.01,0.1,1,10,100,1000},
        xticklabels={$10^{-2}$,$10^{-1}$,$10^0$,$10^1$,$10^2$,$10^3$},
        ymin=0,
        ymax=32,
        xmax=1300,
        bar width=2pt,
        xlabel={Latency (s)},
        ylabel={The Number of Clients},
        legend columns=2,
        legend style={
            at={(0.5,1.4)},
            anchor=north,
            font=\large,
        },
        legend cell align=left,
        legend entries={
            {Read on-chain},
            {Access control verification},
            {Off-chain consensus},
            {Read},
            {ABE decryption}
        },
    ]
    
    \addplot+[draw opacity=0,color=Color03!70] coordinates {(0.0238,0.5) (0.0330,1.5) (0.0328,2.5) (0.0227,3.5) (0.0260,4.5) (0.2119,5.5) (0.2330,6.5) (0.2074,7.5) (0.2104,8.5) (0.2256,9.5) (2.1811,10.5) (2.1904,11.5) (1.9272,12.5) (1.9226,13.5) (1.9518,14.5) (10.7450,15.5) (10.8198,16.5) (9.7079,17.5) (9.5913,18.5) (9.5913,19.5) (21.9831,20.5) (19.3803,21.5) (19.5359,22.5) (20.0533,23.5) (21.7480,24.5) (38.7705,25.5) (43.8221,26.5) (38.7280,27.5) (39.0818,28.5) (38.5460,29.5)}; 
    \addplot+[draw opacity=0,color=Color02!70] coordinates {(0.3864,0.5) (0.3901,1.5) (0.3778,2.5) (0.3825,3.5) (0.3859,4.5) (3.7998,5.5) (3.8741,6.5) (3.6954,7.5) (3.8510,8.5) (3.8212,9.5) (37.7376,10.5) (37.4945,11.5) (38.0846,12.5) (38.4621,13.5) (38.1057,14.5) (190.4470,15.5) (190.6717,16.5) (190.4777,17.5) (190.5428,18.5) (190.8500,19.5) (384.4995,20.5) (383.7209,21.5) (383.5859,22.5) (383.0903,23.5) (381.2003,24.5) (763.9972,25.5) (773.6738,26.5) (770.4529,27.5) (776.2196,28.5) (762.8304,29.5)}; 
    \addplot+[draw opacity=0,color=Color01!70] coordinates {(0.1926,0.5) (0.1587,1.5) (0.1914,2.5) (0.2034,3.5) (0.2061,4.5) (0.5379,5.5) (0.3628,6.5) (0.3014,7.5) (0.3779,8.5) (0.5471,9.5) (1.4264,10.5) (1.0621,11.5) (1.0553,12.5) (0.9585,13.5) (1.0274,14.5) (4.1750,15.5) (4.7265,16.5) (5.0014,17.5) (4.2274,18.5) (4.3411,19.5) (8.1241,20.5) (7.5519,21.5) (8.2365,22.5) (8.0521,23.5) (8.7756,24.5) (14.6056,25.5) (15.5167,26.5) (13.5989,27.5) (15.0585,28.5) (14.6854,29.5)}; 
    \addplot+[draw opacity=0,color=Color04!70] coordinates {(0.0240,0.5) (0.0190,1.5) (0.0200,2.5) (0.0170,3.5) (0.0180,4.5) (0.1586,5.5) (0.1668,6.5) (0.1619,7.5) (0.1726,8.5) (0.1658,9.5) (2.0378,10.5) (1.8874,11.5) (1.8427,12.5) (1.9055,13.5) (1.8822,14.5) (10.3062,15.5) (10.2482,16.5) (10.2564,17.5) (10.1717,18.5) (10.0376,19.5) (19.8561,20.5) (19.9678,21.5) (20.1600,22.5) (20.1104,23.5) (20.0658,24.5) (39.3375,25.5) (40.4630,26.5) (39.1676,27.5) (39.3500,28.5) (41.0771,29.5)};
    \addplot+[draw opacity=0,color=Color06!70] coordinates {(0.2099,0.5) (0.2199,1.5) (0.2119,2.5) (0.2201,3.5) (0.2125,4.5) (2.0542,5.5) (2.0639,6.5) (2.0679,7.5) (2.1056,8.5) (2.0536,9.5) (20.6104,10.5) (20.6019,11.5) (20.6178,12.5) (20.5561,13.5) (20.5457,14.5) (103.0911,15.5) (103.1654,16.5) (103.0232,17.5) (102.8500,18.5) (103.0460,19.5) (206.0212,20.5) (206.0435,21.5) (205.8512,22.5) (205.9709,23.5) (206.6990,24.5) (412.2534,25.5) (412.0741,26.5) (411.8803,27.5) (411.5122,28.5) (412.5165,29.5)};
    
    \end{axis}
\end{tikzpicture}

%% file: evl/UAD-b.tex
\definecolor{Color01}{HTML}{A30141}
\definecolor{Color02}{HTML}{EC6043}
\definecolor{Color03}{HTML}{FCCA78}
\definecolor{Color04}{HTML}{F9FCB7}
\definecolor{Color05}{HTML}{aadda5}
\definecolor{Color06}{HTML}{3893b8}
\definecolor{Color07}{HTML}{4966ae}
  
\begin{tikzpicture}[scale=0.48]
    \begin{axis}[
        xbar stacked,
        axis lines*=left,
        x=0.815cm,
        y=0.12cm,
        axis line style={-latex},
        xmode=log,
        ytick={2.5,7.5,12.5,17.5,22.5,27.5},
        extra y ticks={5,10,15,20,25,30},
        extra y tick labels={,,,,,},
        extra y tick style={major tick length=264pt, thick, black},
        yticklabels={1,10,100,500,1000,2000},
        xtick={0.01,0.1,1,10,100,1000},
        xticklabels={$10^{-2}$,$10^{-1}$,$10^0$,$10^1$,$10^2$,$10^3$},
        ymin=0,
        ymax=32,
        xmax=700,
        bar width=2pt,
        xlabel={Latency (s)},
        ylabel={The Number of Clients},
        legend columns=2,
        legend style={
            at={(0.495,1.4)},
            anchor=north,
            font=\large,
        },
        legend cell align=left,
        legend entries={
            {Read on-chain},
            {Access control verification},
            {Off-chain consensus},
            {Read},
            {BE decryption}
        },
    ]

\addplot+[draw opacity=0,color=Color03!70] coordinates {(0.0289,0.5) (0.0262,1.5) (0.0257,2.5) (0.0353,3.5) (0.0230,4.5) (0.2199,5.5) (0.2054,6.5) (0.2161,7.5) (0.2242,8.5) (0.2154,9.5) (2.1885,10.5) (2.1226,11.5) (2.2934,12.5) (2.1792,13.5) (2.2110,14.5) (10.8224,15.5) (10.8278,16.5) (10.7806,17.5) (10.7871,18.5) (10.8732,19.5) (21.2188,20.5) (21.2462,21.5) (21.1752,22.5) (21.1985,23.5) (21.5518,24.5) (42.7916,25.5) (43.1810,26.5) (44.0577,27.5) (49.8402,28.5) (44.0802,29.5)}; 
\addplot+[draw opacity=0,color=Color02!70] coordinates {(0.1342,0.5) (0.1303,1.5) (0.1381,2.5) (0.1342,3.5) (0.1301,4.5) (1.3002,5.5) (1.3052,6.5) (1.3084,7.5) (1.3154,8.5) (1.3099,9.5) (13.0422,10.5) (13.0229,11.5) (13.0944,12.5) (13.0550,13.5) (12.9951,14.5) (65.2077,15.5) (65.1741,16.5) (65.2462,17.5) (65.0539,18.5) (65.0961,19.5) (130.4620,20.5) (130.3233,21.5) (130.4879,22.5) (130.3406,23.5) (130.3302,24.5) (260.6145,25.5) (293.5809,26.5) (333.7855,27.5) (332.0579,28.5) (328.6577,29.5)}; 
\addplot+[draw opacity=0,color=Color01!70] coordinates {(0.2285,0.5) (0.1965,1.5) (0.1998,2.5) (0.1981,3.5) (0.2120,4.5) (0.3992,5.5) (0.3701,6.5) (0.3196,7.5) (0.4377,8.5) (0.4383,9.5) (1.0389,10.5) (1.2968,11.5) (1.4209,12.5) (1.4143,13.5) (1.0289,14.5) (4.6510,15.5) (5.0602,16.5) (4.6697,17.5) (5.2498,18.5) (3.8966,19.5) (7.6903,20.5) (8.2089,21.5) (8.4038,22.5) (7.5773,23.5) (7.9830,24.5) (14.5854,25.5) (16.6813,26.5) (16.5446,27.5) (15.5107,28.5) (16.0163,29.5)}; 
\addplot+[draw opacity=0,color=Color04!70] coordinates {(0.0640,0.5) (0.0567,1.5) (0.0517,2.5) (0.0527,3.5) (0.0538,4.5) (0.0723,5.5) (0.0662,6.5) (0.0693,7.5) (0.0689,8.5) (0.0656,9.5) (1.0398,10.5) (1.1440,11.5) (1.1690,12.5) (1.1467,13.5) (1.0259,14.5) (10.7007,15.5) (10.7814,16.5) (10.5987,17.5) (10.6587,18.5) (10.1632,19.5) (33.6388,20.5) (33.3464,21.5) (33.0842,22.5) (32.9028,23.5) (32.9921,24.5) (94.0273,25.5) (73.3157,26.5) (92.4975,27.5) (92.8712,28.5) (90.2288,29.5)};
\addplot+[draw opacity=0,color=Color06!70] coordinates {(0.0050,0.5) (0.0040,1.5) (0.0039,2.5) (0.0040,3.5) (0.0038,4.5) (0.0059,5.5) (0.0058,6.5) (0.0060,7.5) (0.0056,8.5) (0.0055,9.5) (0.0237,10.5) (0.0220,11.5) (0.0222,12.5) (0.0220,13.5) (0.0229,14.5) (0.1005,15.5) (0.0971,16.5) (0.0970,17.5) (0.0997,18.5) (0.0990,19.5) (0.1887,20.5) (0.1942,21.5) (0.1986,22.5) (0.1941,23.5) (0.1966,24.5) (0.3831,25.5) (0.3830,26.5) (0.3852,27.5) (0.3673,28.5) (0.3883,29.5)};

\legend{}

\end{axis}
\end{tikzpicture}

%% file: evl/UAD-c.tex
\definecolor{Color01}{HTML}{A30141}
\definecolor{Color02}{HTML}{EC6043}
\definecolor{Color03}{HTML}{FCCA78}
\definecolor{Color04}{HTML}{F9FCB7}
\definecolor{Color05}{HTML}{aadda5}
\definecolor{Color06}{HTML}{3893b8}
\definecolor{Color07}{HTML}{4966ae}

\begin{tikzpicture}[scale=0.48]
    \begin{axis}[
        xbar stacked,
        axis lines*=left,
        x=0.827cm,
        y=0.12cm,
        axis line style={-latex},
        xmode=log,
        ytick={2.5,7.5,12.5,17.5,22.5,27.5},
        extra y ticks={5,10,15,20,25,30},
        extra y tick labels={,,,,,},
        extra y tick style={major tick length=264pt, thick, black},
        yticklabels={1,10,100,500,1000,2000},
        xtick={0.01,0.1,1,10,100,1000},
        xticklabels={$10^{-2}$,$10^{-1}$,$10^0$,$10^1$,$10^2$,$10^3$},
        ymin=0,
        ymax=32,
        xmax=500,
        bar width=2pt,
        xlabel={Latency (s)},
        ylabel={The Number of Clients},
        legend columns=2,
        legend style={
            at={(0.495,1.4)},
            anchor=north,
            font=\large,
        },
        legend cell align=left,
        legend entries={
            {Read on-chain},
            {Access control verification},
            {Off-chain consensus},
            {Read},
            {TE decryption}
        },
    ]

\addplot+[draw opacity=0,color=Color03!70] coordinates {(0.0189,0.5) (0.0229,1.5) (0.0212,2.5) (0.0226,3.5) (0.0219,4.5) (0.2165,5.5) (0.2045,6.5) (0.1843,7.5) (0.1958,8.5) (0.2127,9.5) (2.0276,10.5) (2.1015,11.5) (1.8793,12.5) (2.0800,13.5) (2.1195,14.5) (10.4850,15.5) (9.6720,16.5) (10.4130,17.5) (9.5373,18.5) (9.7074,19.5) (20.9464,20.5) (18.6720,21.5) (20.8515,22.5) (27.8675,23.5) (19.2108,24.5) (42.4408,25.5) (42.6222,26.5) (38.1205,27.5) (38.5969,28.5) (40.9464,29.5)}; 
\addplot+[draw opacity=0,color=Color02!70] coordinates {(0.1442,0.5) (0.1411,1.5) (0.1392,2.5) (0.1381,3.5) (0.1467,4.5) (1.3728,5.5) (1.4105,6.5) (1.3513,7.5) (1.3940,8.5) (1.3847,9.5) (12.9935,10.5) (13.0383,11.5) (13.8579,12.5) (13.0136,13.5) (13.1028,14.5) (64.7847,15.5) (72.8807,16.5) (65.1323,17.5) (74.6786,18.5) (69.0771,19.5) (129.8209,20.5) (149.6716,21.5) (135.9178,22.5) (149.6716,23.5) (139.2035,24.5) (263.0904,25.5) (279.8726,26.5) (306.1763,27.5) (302.4952,28.5) (300.6716,29.5)}; 
\addplot+[draw opacity=0,color=Color01!70] coordinates {(0.1725,0.5) (0.1688,1.5) (0.1878,2.5) (0.2154,3.5) (0.2340,4.5) (0.3735,5.5) (0.3859,6.5) (0.3850,7.5) (0.4076,8.5) (0.5109,9.5) (1.0866,10.5) (1.0921,11.5) (1.3618,12.5) (1.1001,13.5) (0.9582,14.5) (5.7118,15.5) (3.4639,16.5) (4.5493,17.5) (4.4284,18.5) (4.7685,19.5) (7.8941,20.5) (7.3173,21.5) (8.3173,22.5) (8.7620,23.5) (8.5307,24.5) (16.6332,25.5) (15.4857,26.5) (15.3242,27.5) (15.5481,28.5) (19.7342,29.5)}; 
\addplot+[draw opacity=0,color=Color04!70] coordinates {(0.0172,0.5) (0.0181,1.5) (0.0183,2.5) (0.0183,3.5) (0.0194,4.5) (0.1659,5.5) (0.1659,6.5) (0.1628,7.5) (0.1736,8.5) (0.1674,9.5) (1.9948,10.5) (2.0588,11.5) (2.0747,12.5) (1.9850,13.5) (1.9081,14.5) (11.3434,15.5) (10.3012,16.5) (10.7142,17.5) (10.9244,18.5) (11.0582,19.5) (20.5431,20.5) (20.3176,21.5) (20.6104,22.5) (20.7231,23.5) (20.9796,24.5) (41.9365,25.5) (41.0678,26.5) (41.1833,27.5) (41.3797,28.5) (43.5290,29.5)};
\addplot+[draw opacity=0,color=Color06!70] coordinates {(0.0050,0.5) (0.0054,1.5) (0.0050,2.5) (0.0050,3.5) (0.0050,4.5) (0.0138,5.5) (0.0141,6.5) (0.0139,7.5) (0.0152,8.5) (0.0139,9.5) (0.1035,10.5) (0.1051,11.5) (0.1162,12.5) (0.1079,13.5) (0.1049,14.5) (0.5131,15.5) (0.5115,16.5) (0.5106,17.5) (0.5356,18.5) (0.5215,19.5) (1.0220,20.5) (1.0136,21.5) (1.0295,22.5) (1.0261,23.5) (1.0201,24.5) (2.0342,25.5) (2.0483,26.5) (2.0464,27.5) (2.0451,28.5) (2.0552,29.5)};

\legend{}

\end{axis}
\end{tikzpicture}

%% file: tex/conclusion.tex
\section{Conclusion}
FACOS emphasizes the significance of access control-based mechanisms to enhance data privacy and accountability within the permissioned blockchain. Furthermore, we stress the importance of combining BFT off-chain storage solutions with on-chain hashing, especially for managing large datasets. This approach addresses regulatory requirements and simplifies data transfer challenges. To ensure precise permission management and privacy enhancement, we have implemented client-based access control methods tailored to diverse client needs and roles. Additionally, we introduced a trusted verifier to validate and approve client requests. Through comprehensive empirical assessments, our research demonstrates the efficiency and effectiveness of both on-chain and off-chain solutions in addressing the complex challenges of data storage and transfer in sensitive data environments. These findings provide a compelling strategy to navigate the intricacies of data management in today's data-driven landscape.